\documentclass[useAMS,usenatbib]{mn2e}
\usepackage{graphicx}
\usepackage[T1]{fontenc}
\usepackage{color}

\title[EoR Window Recovery]{The Effect of Foreground Mitigation Strategy on EoR Window Recovery}
\author[E. Chapman et al.]
{Emma Chapman,$^1$\thanks{e.chapman@imperial.ac.uk}
Saleem Zaroubi,$^2$ Filipe B. Abdalla,$^3$ Fred Dulwich,$^{4}$ 
\newauthor Vibor Jeli\'{c},$^{2,5,6}$ Benjamin Mort$^{4}$\\ 
$^1$Department of Physics, Blackett Laboratory, Imperial College, London SW7 2AZ, UK \\
$^2$Kapteyn Astronomical Institute, University of Groningen, PO Box 800, 700AV, Groningen, the Netherlands \\
$^3$Department of Physics \& Astronomy, University College London, Gower Street, London, WC1E 6BT \\
$^4$Oxford e-Research Centre, University of Oxford, Keble Rd, Oxford, UK, OX1 3QG \\
$^5$ASTRON - the Netherlands Institute for Radio Astronomy, PO Box
2, 7990 AA Dwingeloo, the Netherlands \\
$^6$Ru{\dj}er Bo\v{s}kovi\'{c} Institute, Bijeni\v{c}ka cesta 54, 10000 Zagreb, Croatia}

\def\LaTeX{L\kern-.36em\raise.3ex\hbox{a}\kern-.15em
    T\kern-.1667em\lower.7ex\hbox{E}\kern-.125emX}

\begin{document}

\maketitle

\begin{abstract}
The removal of the Galactic and extragalactic foregrounds remains a major challenge for those wishing to make a detection of the Epoch of Reionization 21-cm signal. Multiple methods of modelling these foregrounds with varying levels of assumption have been trialled and shown promising recoveries on simulated data. Recently however there has been increased discussion of using the expected shape of the foregrounds in Fourier space to define an EoR window free of foreground contamination. By carrying out analysis within this window only, one can avoid the foregrounds and any statistical bias they might introduce by instead removing these foregrounds. In this paper we discuss the advantages and disadvantages of both foreground removal and foreground avoidance. We create a series of simulations with noise levels in line with both current and future experiments and compare the recovered statistical cosmological signal from foreground avoidance and a simplified, frequency independent foreground removal model. We find that while, for current generation experiments, foreground avoidance enables a better recovery at $k_{perp} > 0.6 \mathrm{Mpc}^{-1}$, foreground removal is able to recover significantly more signal at small $k_{los}$ for both current and future experiments. We also relax the assumption that the foregrounds are smooth by introducing a Gaussian random factor along the line-of-sight and then also spatially. We find that both methods perform well for foreground models with line-of-sight and spatial variations around $0.1\%$ however at levels larger than this foreground removal shows a greater signal recovery.
\end{abstract}


\section{Introduction}

The Epoch of Reionization describes the period of the Universe when uv photons from the first ionizing sources created bubbles in the
neutral hydrogen medium. These bubbles grew and eventually overlapped
to leave the ionized Universe we observe today. Though there exist loose indications of the
end of reionization from the spectra of high redshift quasars \citep[e.g.][]{mortlock11} and integral constraints from the Thomson optical depth \citep{planckcosparam}, this epoch has yet to be directly
detected. Multiple experiments are gathering data
(e.g. Low Frequency Array (LOFAR)\footnote{http://www.lofar.org/} \citep{vanhaarlem13}, Giant Metrewave Radio Telescope
(GMRT)\footnote{http://gmrt.ncra.tifr.res.in/}, Murchison Widefield
Array (MWA)\footnote{http://www.mwatelescope.org/}, Precision Array to
Probe the Epoch of Reionization
(PAPER)\footnote{http://eor.berkeley.edu}) and upper limits from these experiments are pushing ever towards the elusive EoR detection (\citet{Dillon2015,Jacobs2015,Ali2015,Paciga2013,Paciga2011}). The Galactic foregrounds are expected to be three
orders of magnitude larger than the EoR signal for interferometric experiments, providing a
significant challenge for those analysing the incoming
data. Foreground removal methods, where the foregrounds are modelled and removed directly from the data, have developed quickly in the last
few years, moving from specific parametric fits to methods containing
as few assumptions as possible. Alongside these foreground fitting
methods, there has recently been discussion of foreground
avoidance. The supposed form of the foregrounds should restrict them
to a well defined area of the Fourier plane at low $k_{los}$, leaving an `EoR Window',
within which one could carry out analysis of the cosmological
signal free of foreground bias. In this paper we assess the advantages
and disadvantages of foreground removal and foreground avoidance using a
LOFAR-like simulation pipeline, but with consideration to more sensitive future experiments such as the Square Kilometre Array (SKA)\footnote{http://www.skatelescope.org/}. 

Both methods of uncovering the signal have clear motivations. For example
while foreground removal retains information on all modes but
risks contamination of the cosmological signal, foreground avoidance
leaves intact all cosmological signal within the window, but with the
loss of any modes outside this window, which will bias measurements of anisotropic signals such as redshift-space distortions \citet{Jensen2015}. This is a similar situation to
foreground projection methods such as \citep{switzer14} where they
decide between projecting out all the foregrounds and losing cosmological
signal or keeping as much cosmological signal as possible but resigning
oneself to foreground contamination. 

We introduce both approaches,
first foreground avoidance in Sec. \ref{window}, followed by foreground removal in Sec. \ref{fgrem}. We then introduce the simulation pipeline used for
this paper in Sec. \ref{sims} before presenting the cylindrical power spectrum of the two approaches in Sec. \ref{results}. We summarise our main conclusions in Sec. \ref{conc}.

\section{The EoR Window}
\label{window}
 Foreground avoidance was
originally suggested as a way of bypassing the stringent requirements
of foreground subtraction by searching for the signal in a region of
$k$-space where foregrounds are sub-dominant compared to the
signal. A 2D cylindrical power spectrum in $k_{perp}, k_{los}$ shows the areas of
$k$-space where different signal components are dominant and can be used
to define a region where the cosmological signal can be clearly picked
out - an `EoR window'. 

The EoR window is bounded by five physical properties of the
experiment. The $k_{perp}$ boundaries are as a direct result of the
system noise increasing significantly where there is a lack of
baselines. At low $k_{perp}$ the window is bounded by the angular 
extent of the interferometric array which is approximated by the shortest baseline, $L_{min}$. At high $k_{perp}$
the boundary is given by the angular resolution of the instrument, which is effectively the longest baseline 
used in the observation being considered, $L_{max}$. For $k_{los}$ it is the frequency characteristics of the
array which define the boundaries. At low $k_{los}$ it is the bandwidth of
the instrument, $B$, whereas at high $k_{los}$ it is the frequency
resolution of the observation, $\Delta \nu$. These boundaries are defined in Equations \ref{eqn:lims1}-\ref{eqn:lims4},
also described in \citet{vedantham12}, where $D_M(z)$ is the
transverse comoving distance at redshift $z$ and
$E(z) = \sqrt{\Omega_m(1+z)^3 + \Omega_{\Lambda}}$.

\begin{equation}
k_{perp_{max}} = \frac{2 \pi L_{max} \nu_{21}}{c (1+z) D_M(z)}
\label{eqn:lims1}
\end{equation}

\begin{equation}
k_{perp_{min}} = \frac{2 \pi L_{min} \nu_{21}}{c (1+z) D_M(z)}
\label{eqn:lims2}
\end{equation}

\begin{equation}
k_{los_{max}} = \frac{2 \pi H_0 \nu_{21} E(z)}{c (1+z)^2 \Delta \nu}
\label{eqn:lims3}
\end{equation}

\begin{equation}
k_{los_{min}} = \frac{2 \pi H_0 \nu_{21} E(z)}{c (1+z)^2 B}
\label{eqn:lims4}
\end{equation}

\noindent The fifth boundary is a little more complicated.
It has been noted in the literature that the
foregrounds should reside in a specific region in the Fourier domain at low $k_{los}$. Furthermore, a wedge at the high $k_{perp}$ end of this region is defined
due to the mode mixing of the foregrounds as a result of the varying PSF as a
function of frequency. This effect is stronger on larger $k_{perp}$ scales
because of the higher fringe rate associated with long baselines. The wedge was pointed out originally by
\citet{datta10} and can be shown to reach no further than the line \citep[e.g.][]{dillon13} :

\begin{equation}
k_{los} = k_{perp} Sin(\Theta) \frac{H_0 D_c(z) E(z)}{c (1+z)},
\end{equation}

\noindent where $\Theta$ is the field of view and $D_c(z) = \int^z_0 = dz' / E(z')$. 

There has been an excellent series of papers expanding on both the
wedge and the EoR window however there has been no in-depth
comparison between foreground avoidance and a non-parametric
foreground removal method. All the analyses so far have been performed on different simulated or real
data with different levels of complexity and using different methods of
foreground subtraction if any are used at all. 

Originally, \citet{datta10} simulated the EoR window but considered bright point source extragalactic
foregrounds only and performed a simple polynomial
subtraction. 

In \citet{vedantham12}, a careful study of the effect of the PSF
and $uv$ gridding effects on the wedge was performed, however only
bright point sources were considered, without considering
the effects of foreground subtraction. 

\citet{morales12} further defined the window by explaining that
the wedge shape is due to a chromatic instrument response and
information loss in each antenna. They discussed the various mechanisms which can cause power to erroneously enter high $k_{los}$, namely the chromatic instrument response, imperfect foreground models and imperfect instrument calibration.

\citet{trott12} considered
imperfect point source subtraction and the effect on the EoR window,
suggesting that the contamination from residual point sources would
not be a limiting effect in the EoR detection. 

\citet{parsons12} considered diffuse
synchrotron emission alongside extragalactic foregrounds however
they did so assuming the spectral distribution of the synchrotron is a simple
scaling of a $\nu ^{-2.5}$ power law derived from the low resolution Haslam map. This neglects
small-scale power and any potential variation of the spectral
index. 

\citet{pober13} discusses observations with PAPER and notes that the foregrounds extend beyond the expected theoretical limit in $k$-space due to the spectral structure of the foregrounds. They also conclude that as the bulk of the emission contaminating the EoR window is diffuse it is this which is the biggest challenge to deal with as opposed to, say, the point sources.

\citet{dillon13}, used real MWA data to consider different estimators in the power spectrum estimation framework and showed that the frequency dependence of the wedge was in line with theoretical expectations, i.e. that it got brighter and larger in area with decreasing frequency. Using foreground avoidance they extracted upper limits on the EoR power spectrum.

\citet{hazelton13} consider the effect of mode-mixing from non-identical baselines, concluding that, compared to the single baseline mode-mixing usually discussed, power could easily be thrown from the wedge into the window. 

\citet{liu14a,liu14b} produced a comprehensive study of the mathematical formalism in which to describe the wedge and the associated errors, providing a comprehensive framework where one could easily see the probing of finer spatial scales at higher frequencies by any given baseline. They also discuss hoe to maximise the cleanliness of the EoR window using various methods such as different estimators in their power spectrum framework.

\citet{Thyagarajan2015} simulated full-sky instrument and foreground models to show that there is significant contamination of the EoR window from foreground emission outside the primary field of view. They went on to confirm a `pitchfork' structure within the MWA foreground wedge structure, with maxima both relating to the primary beam and the horizon limit \citep{Thyagarajan2015a}.

Here we use a full diffuse foreground model to assess the feasibility
of the EoR window and how the recovered statistical information compares to that recovered using foreground removal. We also model different types of contamination into the foreground signal which compromises the assumption of smoothness along the line-of-sight, and therefore the assumption that the foregrounds will reside in a well-defined area at low $k_{los}$.

\section{Foreground Removal}
\label{fgrem}
Though the foregrounds are expected to be approximately three orders of
magnitudes larger than the cosmological signal for interferometric
data, the two signals have a markedly different frequency structure. While
the cosmological signal is expected to decorrelate on frequency widths on the order of MHz, the foregrounds are expected to be smooth
in frequency. 

The vast majority of foreground removal
methods exploit this smoothness to carry out `line-of-sight' fits to the
foregrounds which can then be subtracted from the data. While early methods assumed this smoothness
implicitly by directly fitting polynomials to the data (e.g. \citealt{santos05}; \citealt{wang06}; \citealt{mcquinn06}; \citealt{bowman06}; \citealt{jelic08}; \citealt{gleser08}; \citealt*{liu09a}; \citealt{liu09b}; \citealt{petrovic11}), there
has recently been increased focus on so-called `blind', or
non-parametric, methods. This is due to the fact that the foregrounds
have never been observed at the frequencies and resolution of the
current experiments and models therefore rely heavily on extrapolation from
low resolution and high frequency maps. Also, the instrumental effect
on the observed foregrounds is by no means likely to be smooth. For
example, the leakage of polarized foregrounds \citep{bernardi10,jelic10,bernardi13,jelic14,asad15} is a serious concern and
would create a decorrelation along the line of sight. While parametric
methods will be unable to model this, methods which make fewer
assumptions about the exact form of the foregrounds have a chance of
modeling non-smooth components. 

Non-parametric methods attempt to avoid assuming any specific
form for the foregrounds and instead use the data to define the
foreground model. For example, `Wp' smoothing as applied to EoR simulations by \citet{harker09b,harker10}, fits
a function along the line of sight according to the data and not some
prior model, penalizing changes in curvature along the line of sight. While
still assuming general smoothness of the foregrounds for the method to be well-motivated, this method includes a smoothing parameter which allows
the user to control how harsh this smoothing condition is to allow for
deviations from the smoothness prior. 

Other non-parametric methods both use a statistical framework known as the mixing model, Eqn. \ref{xas}. This posits that the
foregrounds can be described by a number of different
components which are combined in different ratios according to the
frequency of observation. It should be noted
that these foreground components are not necessarily, or even likely
to be, the separate foreground contributions such as Galactic
free-free and Galactic synchrotron, but instead combinations of them. FastICA as applied to EoR simulations by \citet{chapman12} is
an independent component analysis technique which assumes the
foreground components are statistically independent in order
to model them.  In this paper we will utilise the component
separation method Generalized Morphological Component Analysis (GMCA) \citep{bobin08b}, which we previously successfully applied to
simulated LOFAR data \citep{Ghosh2015,chapman13}. Other component analysis methods can be seen in the literature, for example, \citet{Zhang2015,Bonaldi2015}.

\subsection{GMCA}
GMCA assumes that there
exists a basis in which the foreground components can be termed
sparse, i.e. represented by very few basis coefficients. As the
components are unlikely to have the same few coefficients, the
components can be more easily separated in that basis, in this case
the wavelet basis. GMCA is a blind
source separation (BSS) technique,
such that both the foreground components and the mixing matrix must be
estimated. The reason that GMCA is able to clean the foregrounds so effectively is due to the completely different scale information contained within the foreground signal compared to the cosmological signal and instrumental noise. This leads to very different basis coefficients and the reduction of the cosmological signal and instrumental noise to a `residual' signal. GMCA is not currently able to separate out the cosmological signal alone due to the overwhelmingly small signal-to-noise of the problem. Thus the residual must be post-processed in order to account for the instrumental noise and fully reveal the cosmological signal. For example, in power spectra, the power spectrum of the cosmological signal is revealed by subtracting the known power spectrum of the noise from the power spectrum of the GMCA residual. We now present a brief mathematical framework for GMCA:

Consider an observation of $m$ frequencies each constituting maps of $t$ pixels. The mixing model is as follows:

\begin{equation}
\label{xas}
\mathbfss{X}=\mathbfss{A}\mathbfss{S}+\mathbfss{N}
\end{equation}

\noindent where $\mathbfss{X}$ is the $m \times t$ matrix representing the observed data, $n$ is the number of components to be estimated, $\mathbfss{S}$ is the signal $n \times t$ matrix to be determined, $\mathbfss{A}$ is the $m \times n$ mixing matrix and $\mathbfss{N}$ is the $m \times t$ noise matrix.

We need to estimate both $\mathbfss{S}$ and $\mathbfss{A}$. We aim to find the 21-cm signal as a residual in the separation process, therefore $\mathbfss{S}$ represents the foreground signal and, due to the extremely low signal-to-noise of this problem, the 21-cm signal can be thought of as an insignificant part of the noise.

We can choose to expand $\mathbfss{S}$ in a wavelet basis with the objective of GMCA being to seek an unmixing scheme, through the estimation of $\mathbfss{A}$,
which yields the sparsest components $\mathbfss{S}$ in the wavelet domain.

For more technical details about GMCA, we refer the interested reader to \citealt{bobin07}, \citealt{bobin08a}, \citealt*{bobin08b}, \citealt{bobin12}, where it is shown that sparsity, as used in GMCA, allows for a more precise estimation of the mixing matrix $\mathbfss{A}$ and more robustness to noise than ICA-based techniques.

Though
GMCA is labeled non-parametric due to the lack of specific model for
foregrounds, one must specify
the number of components in the foreground model. This is not a
trivial choice as too small a number and the foregrounds will not be
accurately modelled resulting in large foreground leakage into the
recovered cosmological signal. Too large a number and you risk the leakage of cosmological
signal into the reconstructed foregrounds. In principle, one might attempt to
iterate to the `correct' number of components by minimizing the
cross-correlation coefficient between the residuals and reconstructed
foregrounds cube, however this remains, in some form, a parameter in a
so-called non-parametric method. Here we assume 4 components which we
have previously shown to be suited the foreground simulations we use \citep{chapman13}.

\section{Simulations}
\label{sims}

In order to model the cosmological signal we use the semi-numeric reionization code \textsc{simfast21}\footnote{http://www.simfast21.org/}. The real space cosmological brightness temperature boxes were converted to
an observation cube evolving along the frequency axis using a standard light cone prescription as described in, for example, \citet{datta11}.

The foregrounds are simulated to
consist of Galactic synchrotron, Galactic free-free and unresolved
extragalactic foregrounds according to \citet{jelic08,jelic10} the details of
which we repeat from \citet{chapman12}:

\begin{enumerate}
\item Galactic diffuse synchrotron emission (GDSE) originating from the interaction of free electrons with the Galactic magnetic field. Incorporates both the spatial and frequency variation of $\beta$ by simulating in 3 spatial and 1 frequency dimension before integrating over the $z$-coordinate to get a series of frequency maps. Each line of sight has a slightly different power law. 
\item Galactic localised synchrotron emission originating from supernovae remnants (SNRs). Together with the GDSE, this emission makes up 70 per cent of the total foreground contamination. Two SNRs were randomly placed as discs per 5$^{\circ}$ observing window, with properties such as power law index chosen randomly from the \citet{green06} catalog\footnote{http://www.mrao.cam.ac.uk/surveys/snrs/}.
\item Galactic diffuse free-free emission due to bremsstrahlung radiation in diffuse ionised Galactic gas. This emission contributes only 1 per cent of total foreground contamination, however it still dominates the 21-cm signal. The same method as used for the GDSE is used to obtain maps, however the value of $\beta$ is fixed to -2.15 across the map.
\item Extragalactic foregrounds consisting of contributions from radio galaxies and radio clusters and contributing 27 per cent of the total foreground contamination. The simulated radio galaxies assume a power law and are clustered using a random walk algorithm. The radio clusters have steep power spectra and are based on a cluster catalogue from the Virgo consortium\footnote{http://www.virgo.dur.ac.uk/} and observed mass-luminosity and X-ray-radio luminosity relations.
\end{enumerate} 

\begin{figure}
\includegraphics[width=80mm]{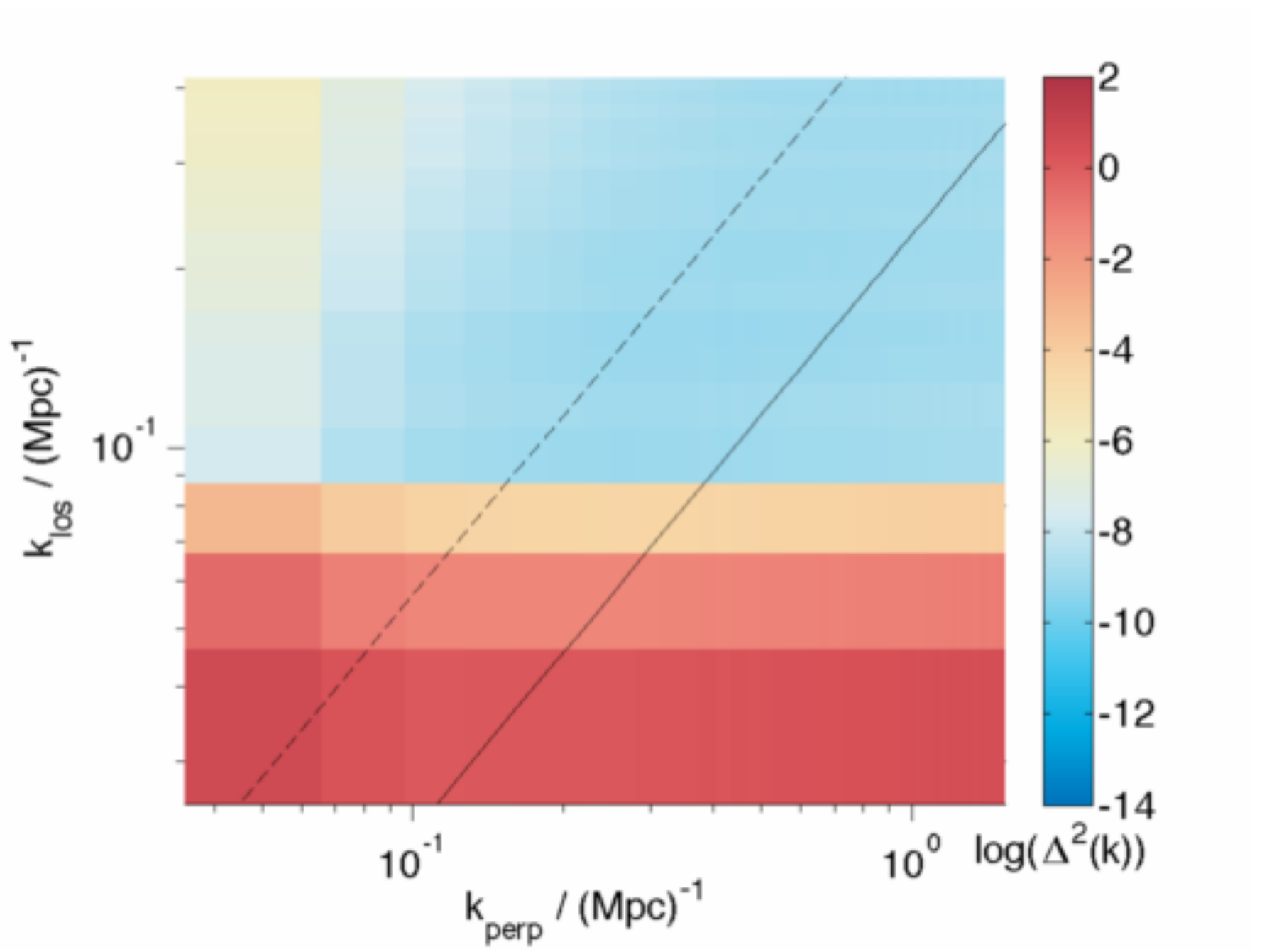}
\includegraphics[width=80mm]{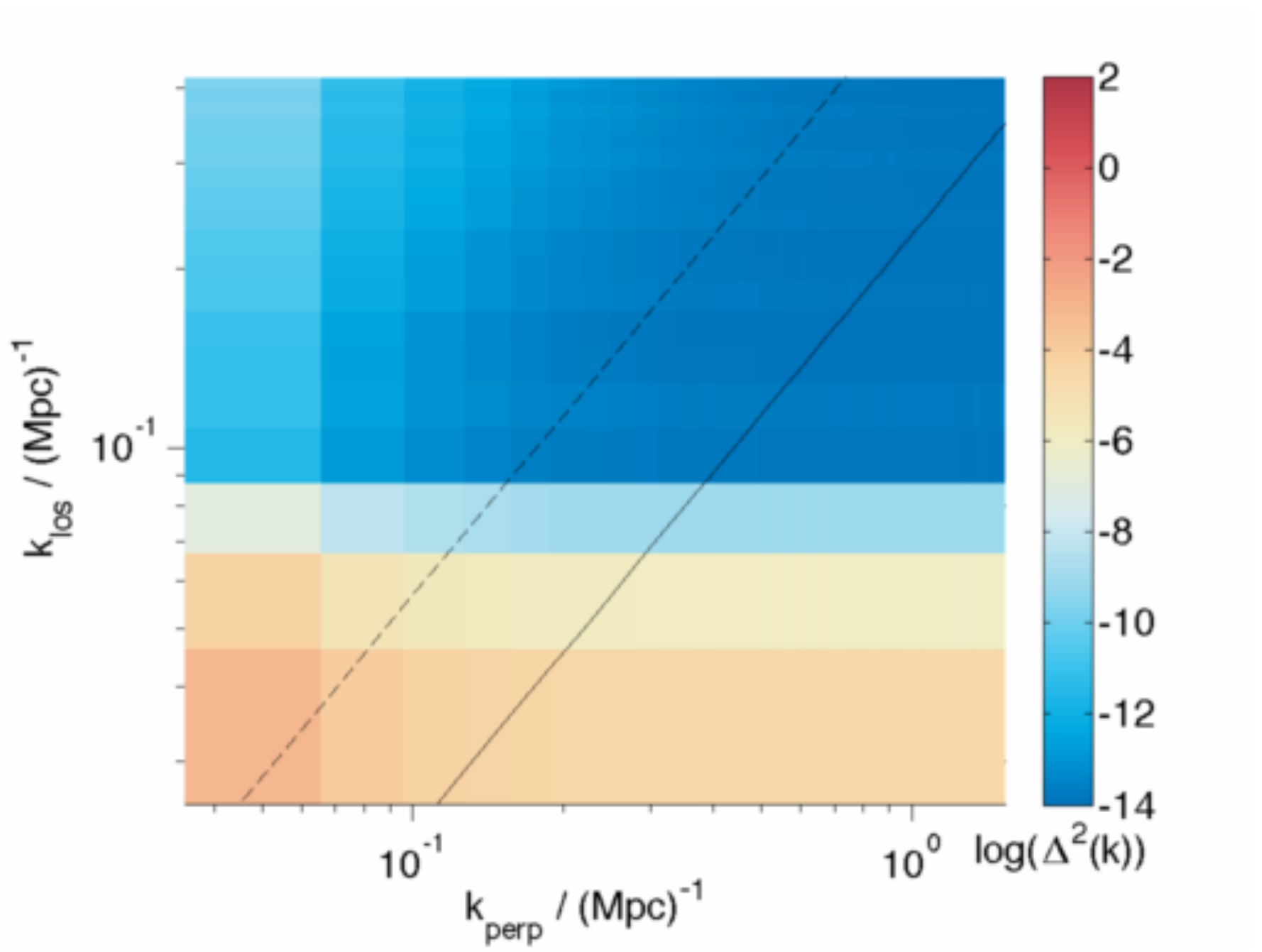}
\includegraphics[width=80mm]{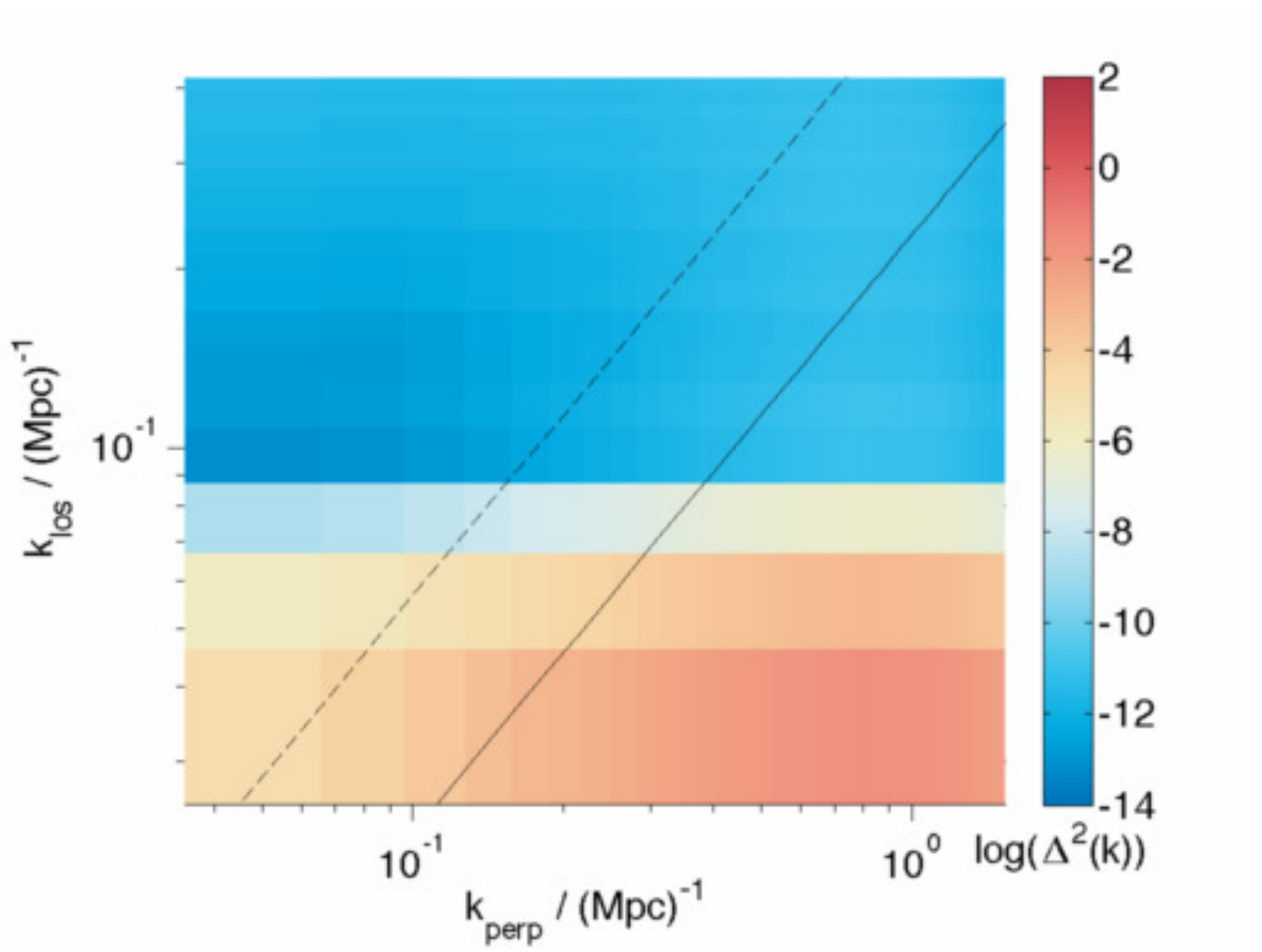}
\caption{The cylindrical power spectrum for the Galactic synchrotron,
  Galactic free-free and extragalactic foregrounds (from top to
  bottom). This is for a 10 MHz bandwidth centered at 165 MHz. The theoretical wedge limit line is shown in black.}
\label{fgcomp_win}
\end{figure}

\begin{figure}
\includegraphics[width=80mm]{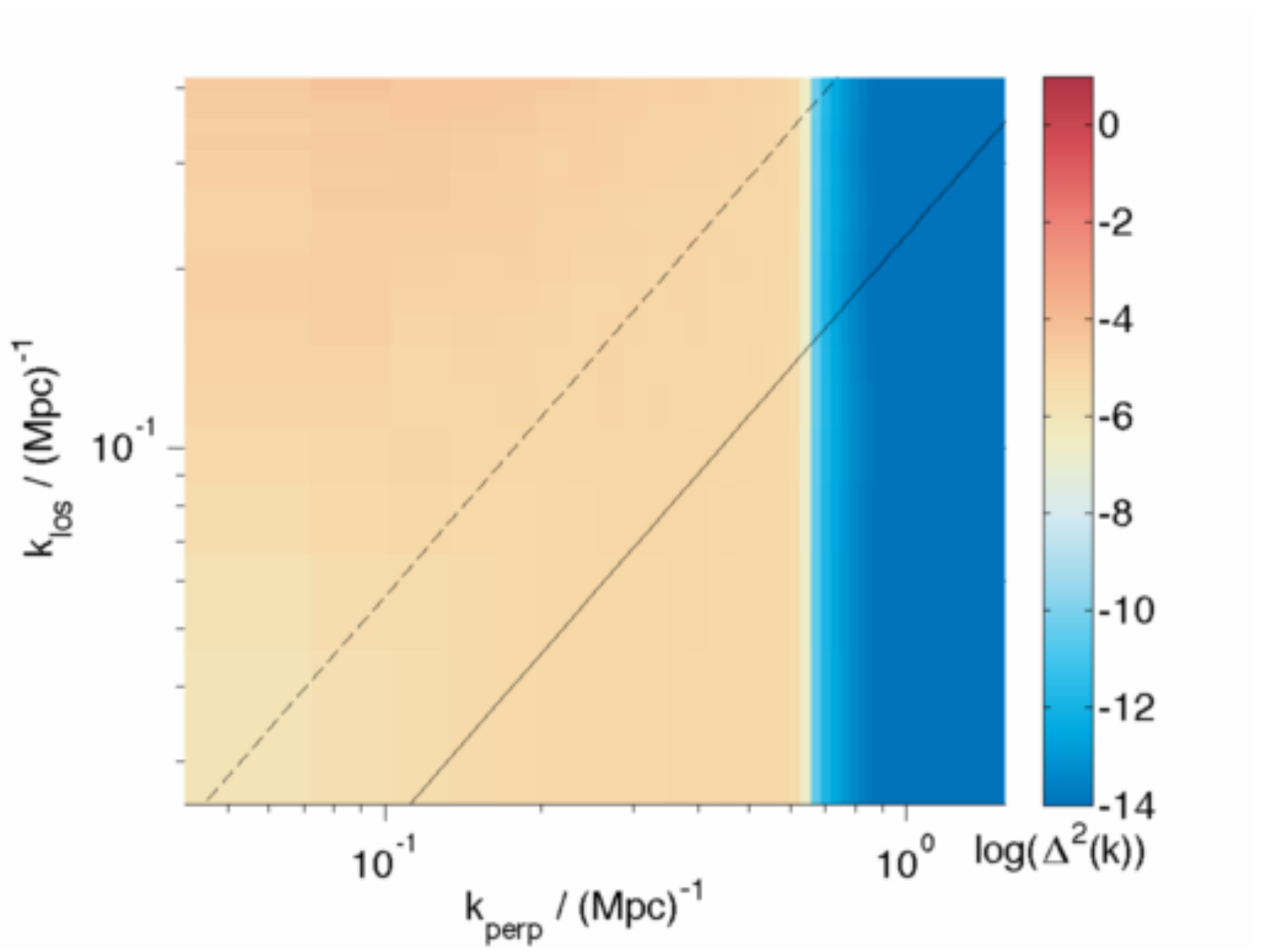}
\includegraphics[width=80mm]{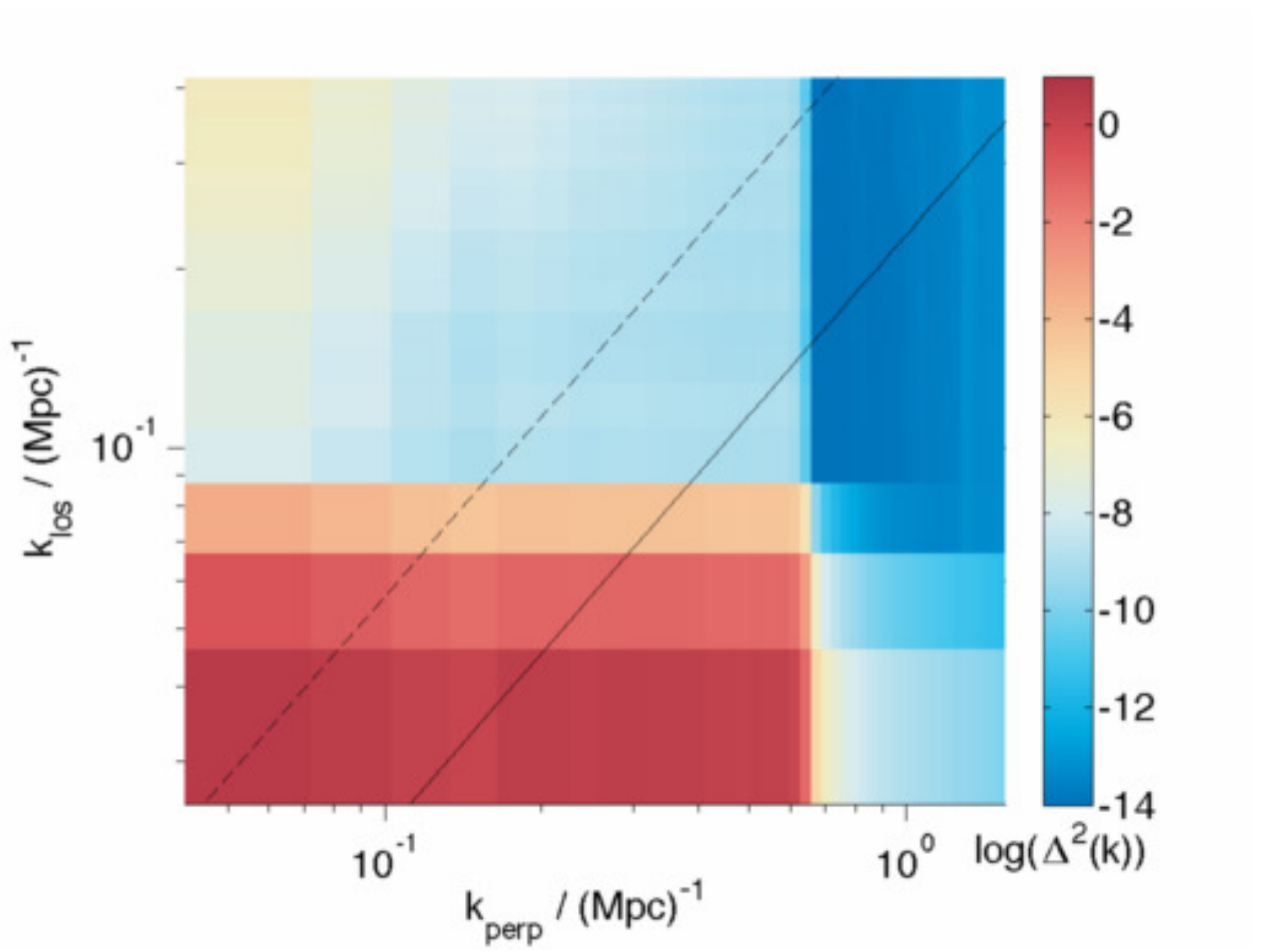}
\caption{The cylindrical power spectrum for the cosmological signal (top)
  and combined foregrounds (bottom) (i.e. sum of Galactic synchrotron,
  Galactic free-free and extragalactic foregrounds) both convolved
  with a LOFAR 115 MHz PSF.}
\label{csfg_win}
\end{figure}

We assume that bright, resolvable, point sources have
been removed accurately and will not limit the signal detection, as
concluded in the literature \citep{trott12}. 

We simulate realistic instrumental effects by using the measurement equation software OSKAR\footnote{http://www.oerc.ox.ac.uk/~ska/oskar2/}. The clean cosmological signal and foregrounds are sampled in UV space using a LOFAR core antenna table, including the primary beam. We choose here to simply sample each slice of the clean cosmological signal and foreground as if it were being observed at 115 MHz. In other words, we are ensuring the resolution of the data is identical throughout the frequency range. It is the authors' findings that current methods within the observation pipeline result in the comparison here being valid. With real data, there would be an upper and lower cut in the UV plane which, in the case where the UV plane is fully and uniformly sampled, will ensure a common resolution. Of course, the UV plane is not always fully and uniformly sampled and hence even this UV cut can result in frequency dependent resolution when, for example, uniform weighting is used. However an alternative, adaptive weighting scheme where by the weighting is optimised to minimise the differential PSF results in a much more well-behaved PSF with frequency (see \citet{2014MNRAS.444..790Y} Fig. 9 and 10). It is possible that the frequency dependence of the instrument can have a non-negligible effect even after the mitigation, in which case the comparison here could be treated as an optimistic case with respects to foreground removal.

The visibilities are then imaged using CASA with uniform weighting to produce 'dirty' images. The spatially correlated instrumental noise is created using OSKAR by filling an empty measurement set with random numbers, imaging and normalising the rms of the image according to the standard noise sensitivity prescription (\citet{thompson01}). Note that different weighting schemes affect the noise estimate via. the system efficiency parameter. In LOFAR calculations we estimate thermal noise
directly from observations, i.e., uniformly weighted Stokes V images and produce a system efficiency from this such that our noise estimates are conservative and realistic. In order to ensure the primary beam does not effect the foreground removal process, we take only the central 4 degrees of the resulting 10 degree images. 

We note that in this work we do not see a wedge structure due to the incomplete treatment of instrumental frequency dependence. This will be addressed in future work. We have however included two theoretical lines relating to the wedge limit on all figures - a dashed line which assumes sources in the entire field of view (in our case 10 degrees) will leak into the main field and an optimistic solid line which assumes only sources within the chosen central 4 degrees (approximately the FWHM of the station beam) will affect the image. Further work on the contamination within the wedge is ongoing in the literature and as such recovery below the theoretical wedge line should be treated with caution.

We refer to the combination of the noise, dirty foregrounds and dirty cosmological
signal as the `total input signal'. We refer to the foreground model estimated by GMCA as the `reconstructed
foregrounds' and the difference between the reconstructed foregrounds
and the total input signal as the `residuals', which will contain the
cosmological signal, noise and any foreground fitting errors. The foreground fitting errors can be defined as the difference between the simulated foregrounds and the reconstructed foregrounds and will contain any cosmological signal or noise which has been fitted as foregrounds as well as an absence of any foregrounds not modelled.

\section{Results}
\label{results}

Unless otherwise stated we present results for cylindrical power spectrum calculated
over a 10 MHz bandwidth segment centered at 165 MHz, where the variance of the
cosmological signal peaks in our simulation. This bandwidth is formed using an extended Blackmann-Nuttall window, though other windows are possible, as discussed in section \ref{sec:win_comp}.

We 3D Fourier transform the data segment and bin the data according to the values of
$k_{perp},k_{los}$. The power at any particular $k_{los}, k_{perp}$,
$P(k_{los},k_{perp}) = \langle\delta(k_{los},k_{perp}) \delta^*(k_{los},k_{perp})\rangle$ is
the average power of all the $uv$ cells in the bin centering on $k_{los},k_{perp}$. We choose to plot the ``dimensionless" power spectrum, $\Delta^2(k_{los,k_{perp}})$, which has units of mK$^2$ throughout this paper:

\begin{equation}
\Delta^2(k_{los},k_{perp}) = P(k_{los},k_{perp}) V \frac{|k|^3}{2\pi^2}
\end{equation}
\noindent where V is the volume of the 10 MHz bandwidth in Mpc.

We first construct cylindrical power spectrum for the different signal
contributions in order to understand how the EoR window is
constructed. In Fig. \ref{fgcomp_win} we see the Galactic synchrotron, Galactic free-free and unresolved
extragalactic foregrounds while in Fig. \ref{csfg_win} we see the dirty
cosmological signal and dirty foregrounds (with all three foreground
contributions summed). The action of the
PSF can be clearly seen as a loss of
power on scales above this (i.e. for $k_{perp}> 0.65 \mathrm{Mpc}^{-1}$). Note that, since we are seeing power spectrum
over a 10 MHz bandwidth only we do not see the evolution of the
cosmological signal over redshift. We see clearly the area of
$k$-space where the foregrounds appear to be at their strongest is at
low $k_{los}$,
however, it should be noted that there is appreciable foreground
contamination across a large proportion of the plane. The cosmological signal has clear contours with the strongest signal at low $k$ -  right beneath the foreground contamination. By choosing to avoid the foregrounds we will implicitly lose information on the strongest part of the cosmological signal. 

\subsection{Choice of Window Function}
\label{sec:win_comp}
\begin{figure}
\begin{minipage}{70mm}
\begin{center}
\includegraphics[width=80mm]{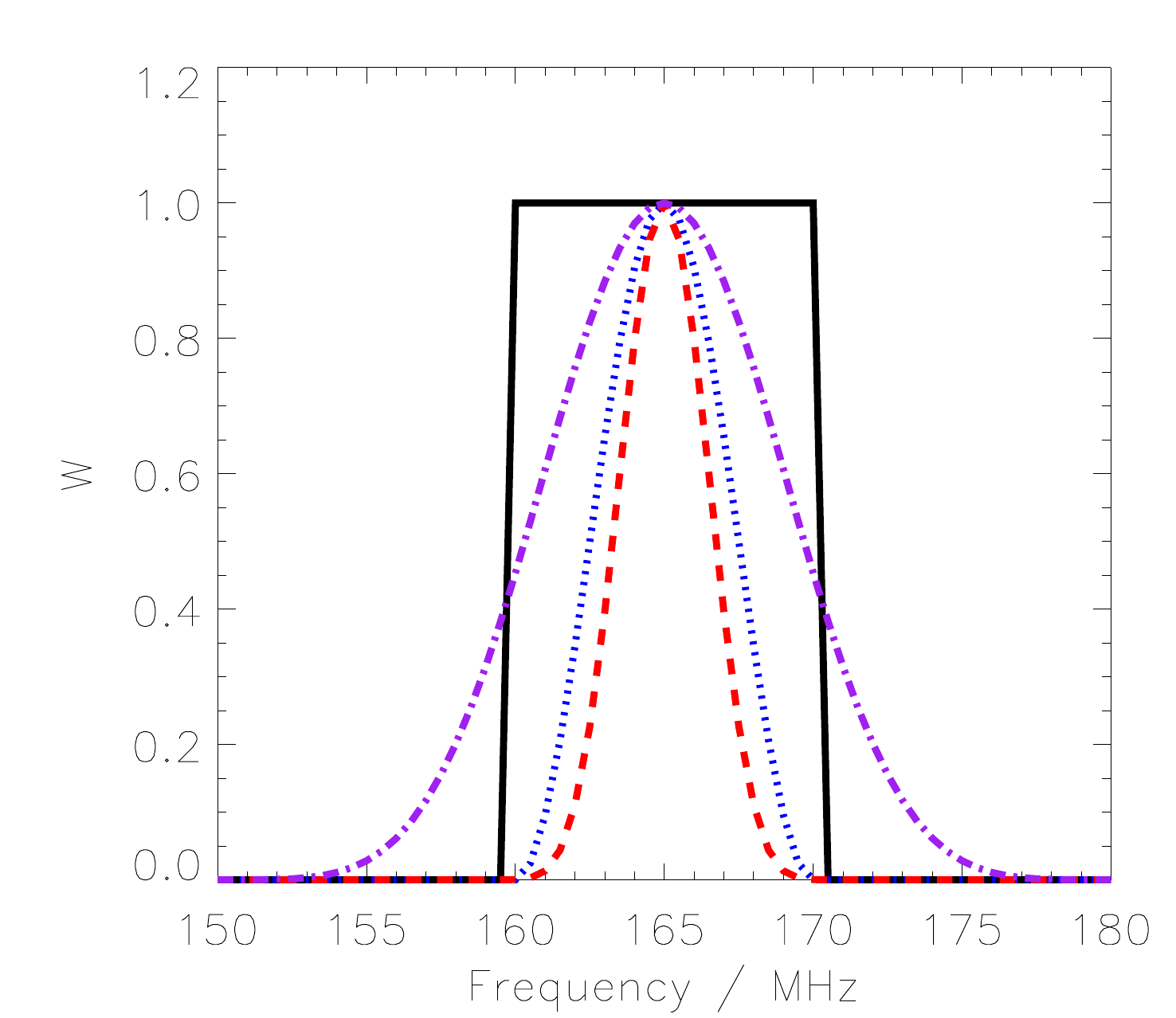}
\caption{The 10 MHz rectangular (black, solid), Hanning (blue, dot), Blackmann-Nuttall (red, dash) and extended Blackman-Nuttall (purple, dash-dot) window.}
\label{window_funcs}
\end{center}
\end{minipage}
\end{figure}

\begin{figure*}
\begin{minipage}{170mm}
\begin{center}
\includegraphics[width=80mm]{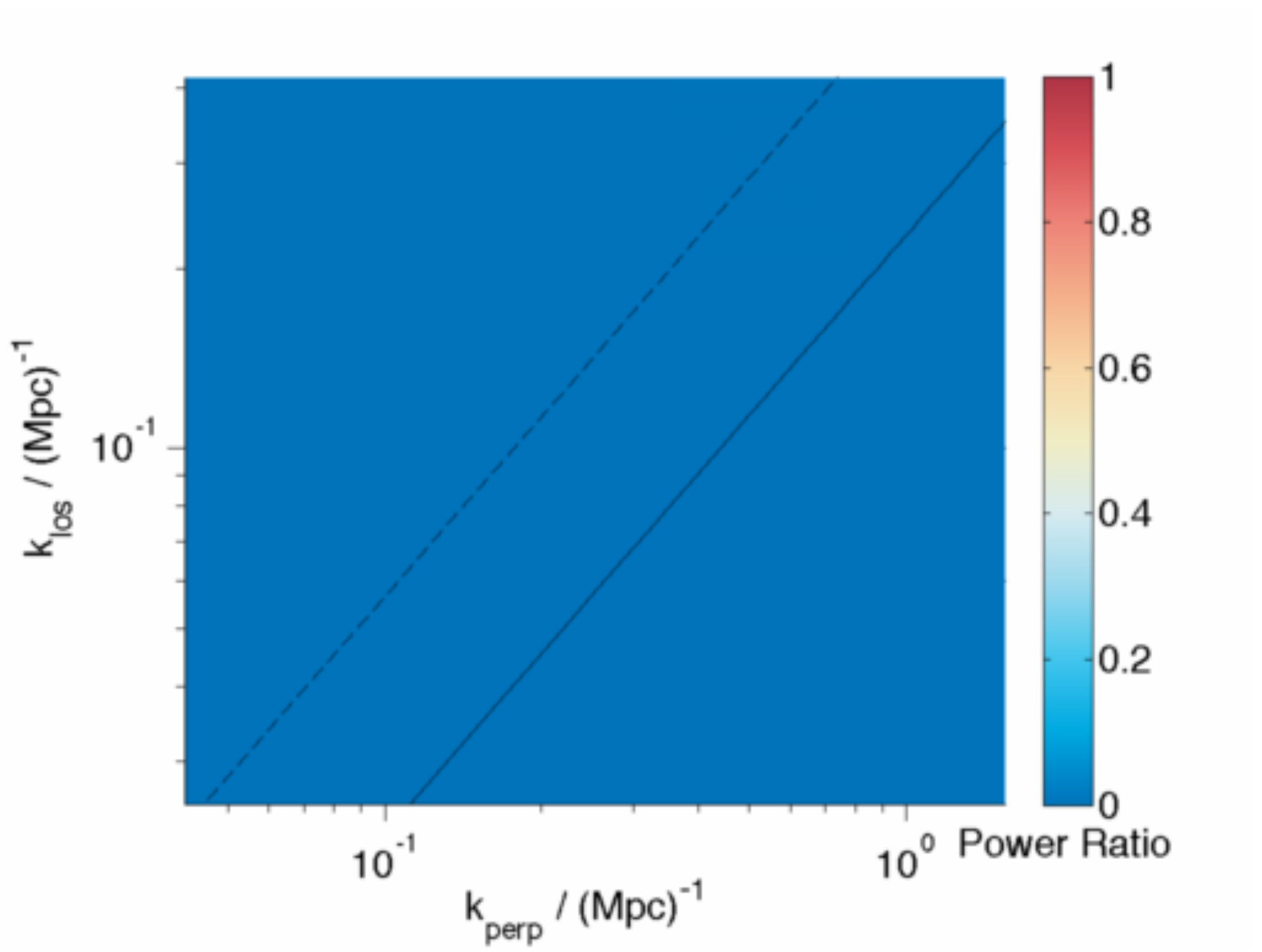}
\includegraphics[width=80mm]{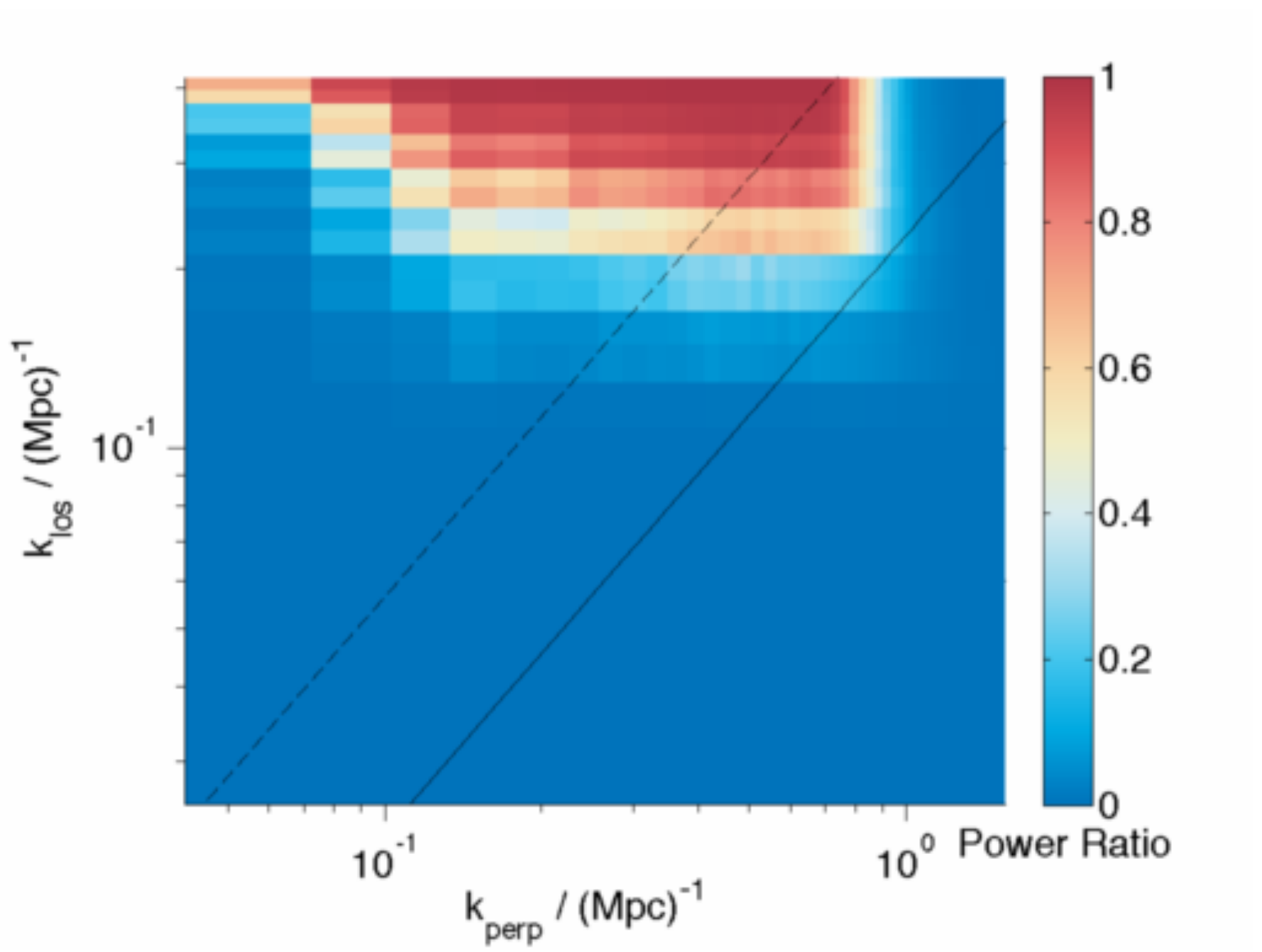}
\includegraphics[width=80mm]{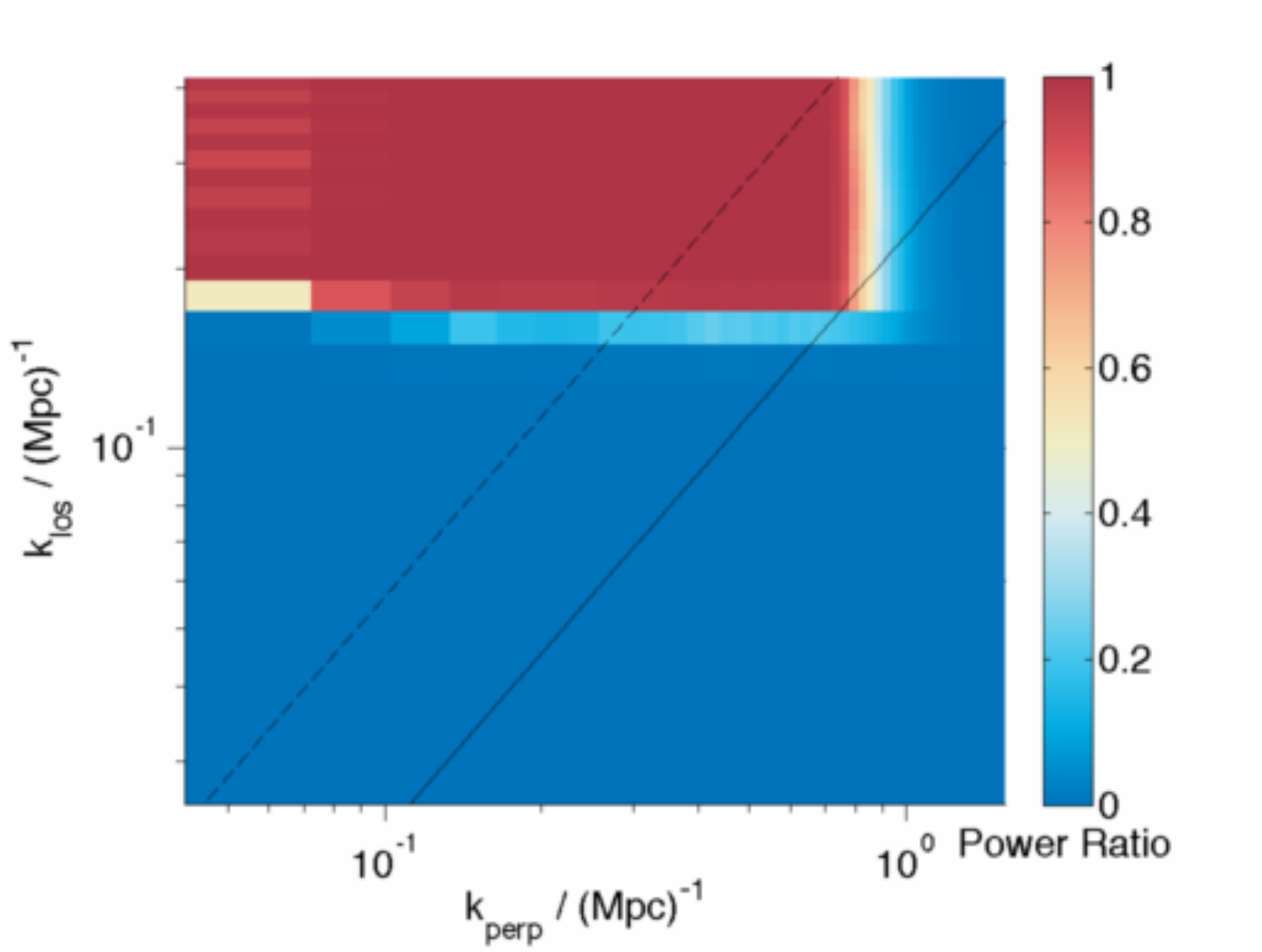}
\includegraphics[width=80mm]{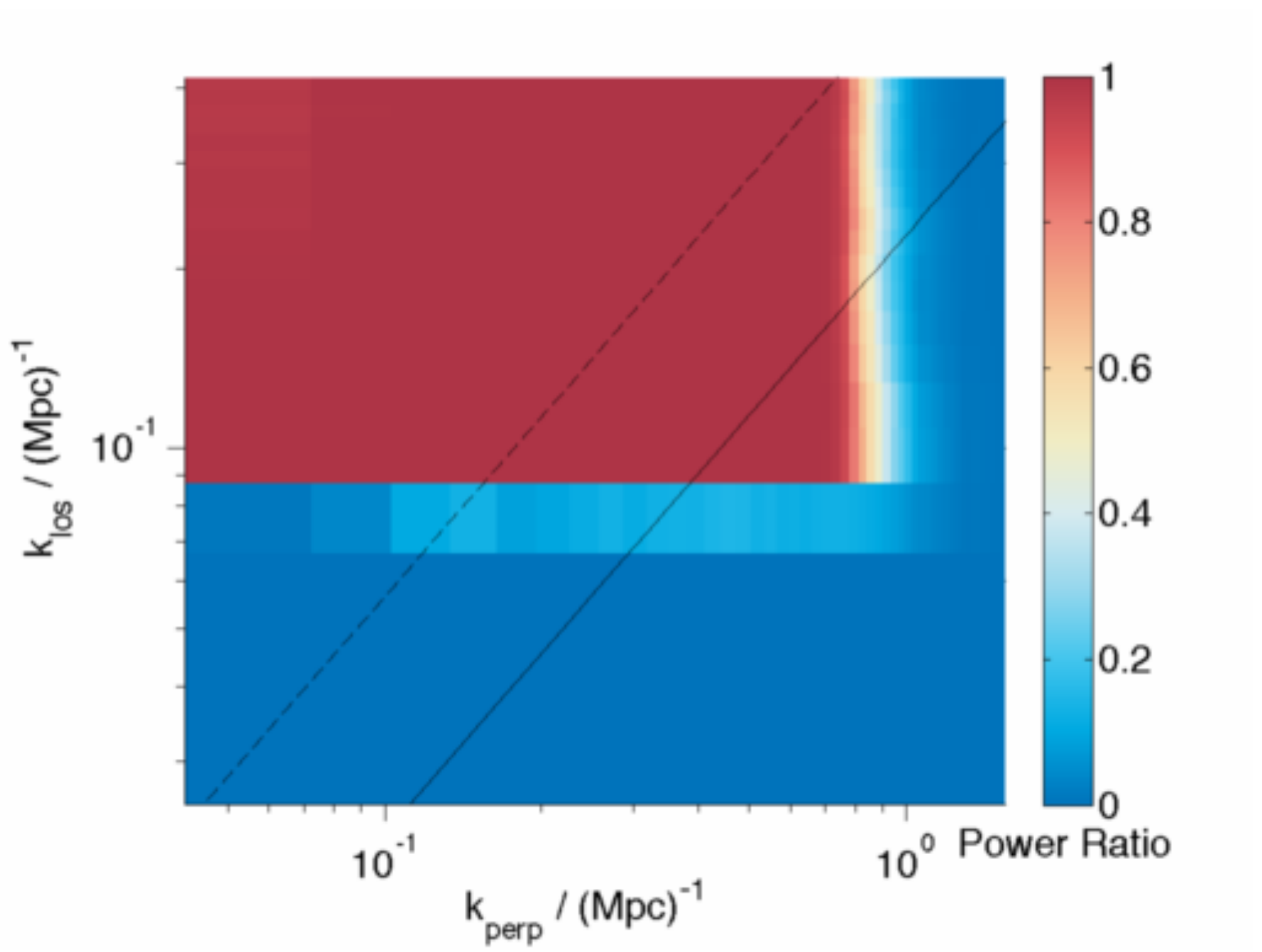}
\caption{The ratio of $\mathrm{cs}/\mathrm{cs+fg}$ for a 10 MHz rectangular, Hanning, Blackmann-Nuttall and extended Blackman-Nuttall window (in reading order). The extended Blackmann-Nuttall provides the largest clean window. The lines represent theoretical wedge contamination limits for sources contributing from within the field of view of 10 degrees (dashed) and from within the FWHM of the station beam (dotted))}
\label{win_comp}
\end{center}
\end{minipage}
\end{figure*}

The cylindrical power
spectrum are calculated by first applying a window function along the
line-of-sight in order to suppress the sidelobes associated with the
Fourier transformation of a finite signal. The choice of this window is not unimportant as different windows can cause different levels of foreground suppression to be balanced with less sensitivity \citep{thyagarajan13,vedantham12}. We compare the cleanliness of the EoR window for four different types of window, rectangular, Hanning, Blackmann-Nuttal and extended Blackman Nuttall (the forms of which are shown in Fig.\ref{window_funcs}) in Fig. \ref{win_comp}.

It is clear that different windows result in different suppression of the foreground contamination of the EoR window. The extended Blackman-Nuttall window seems to result in the most clean window and for all results following we choose to use the extended Blackmann-Nuttall window.

\subsection{Smooth Foreground Models}
\begin{figure*}
\begin{minipage}{170mm}
\begin{center}
\includegraphics[width=80mm]{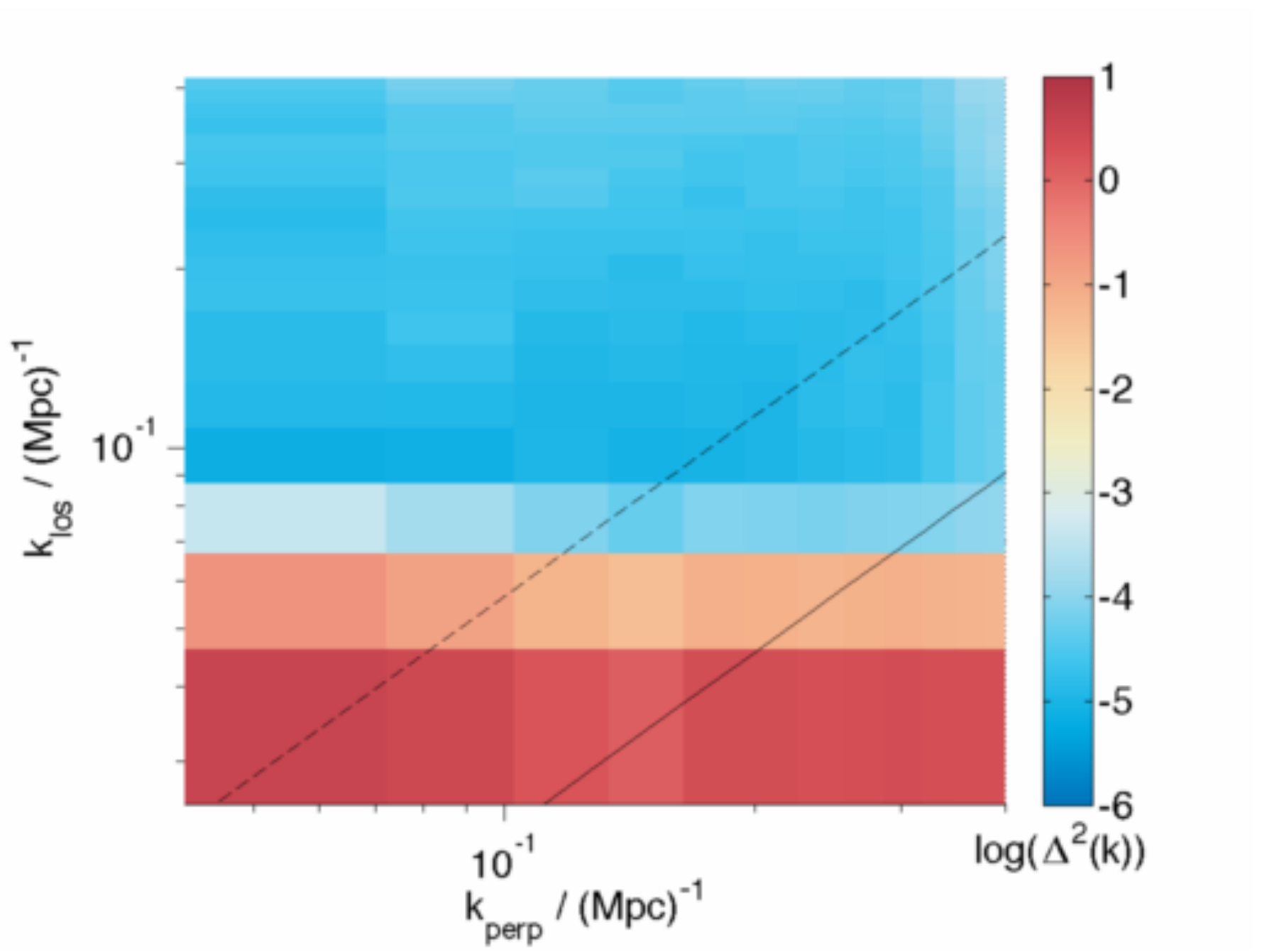}
\includegraphics[width=80mm]{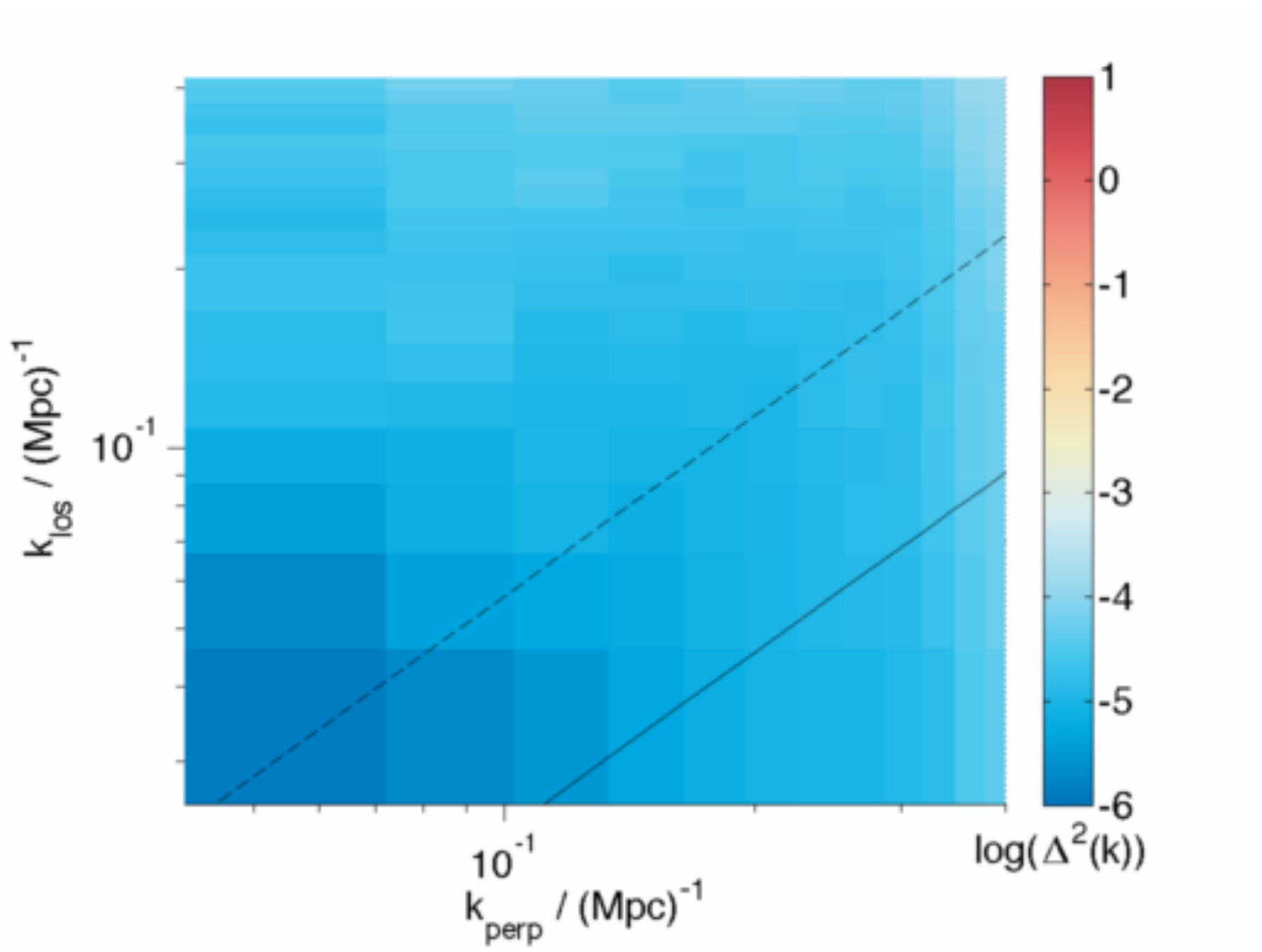}
\includegraphics[width=80mm]{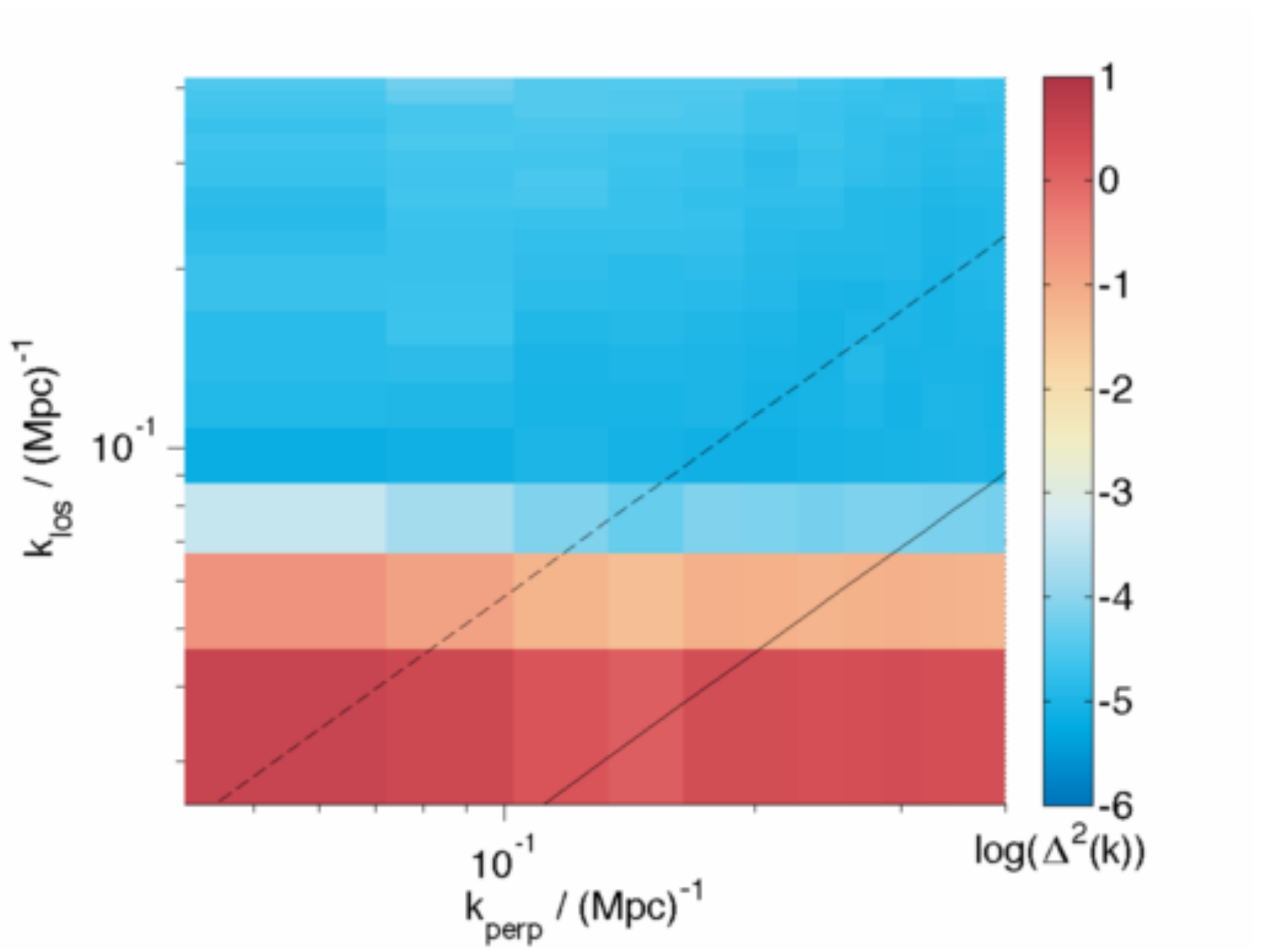}
\includegraphics[width=80mm]{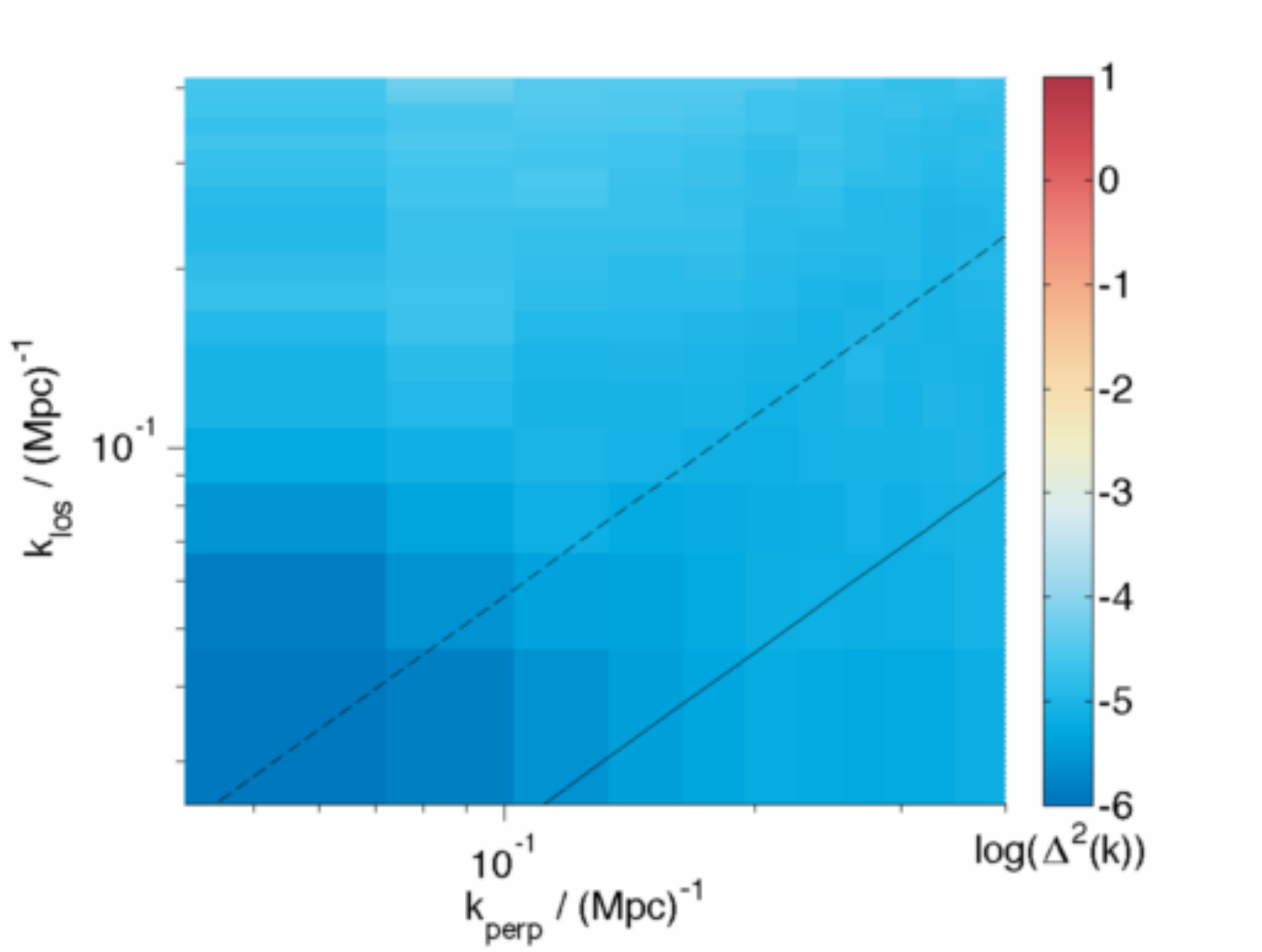}
\includegraphics[width=80mm]{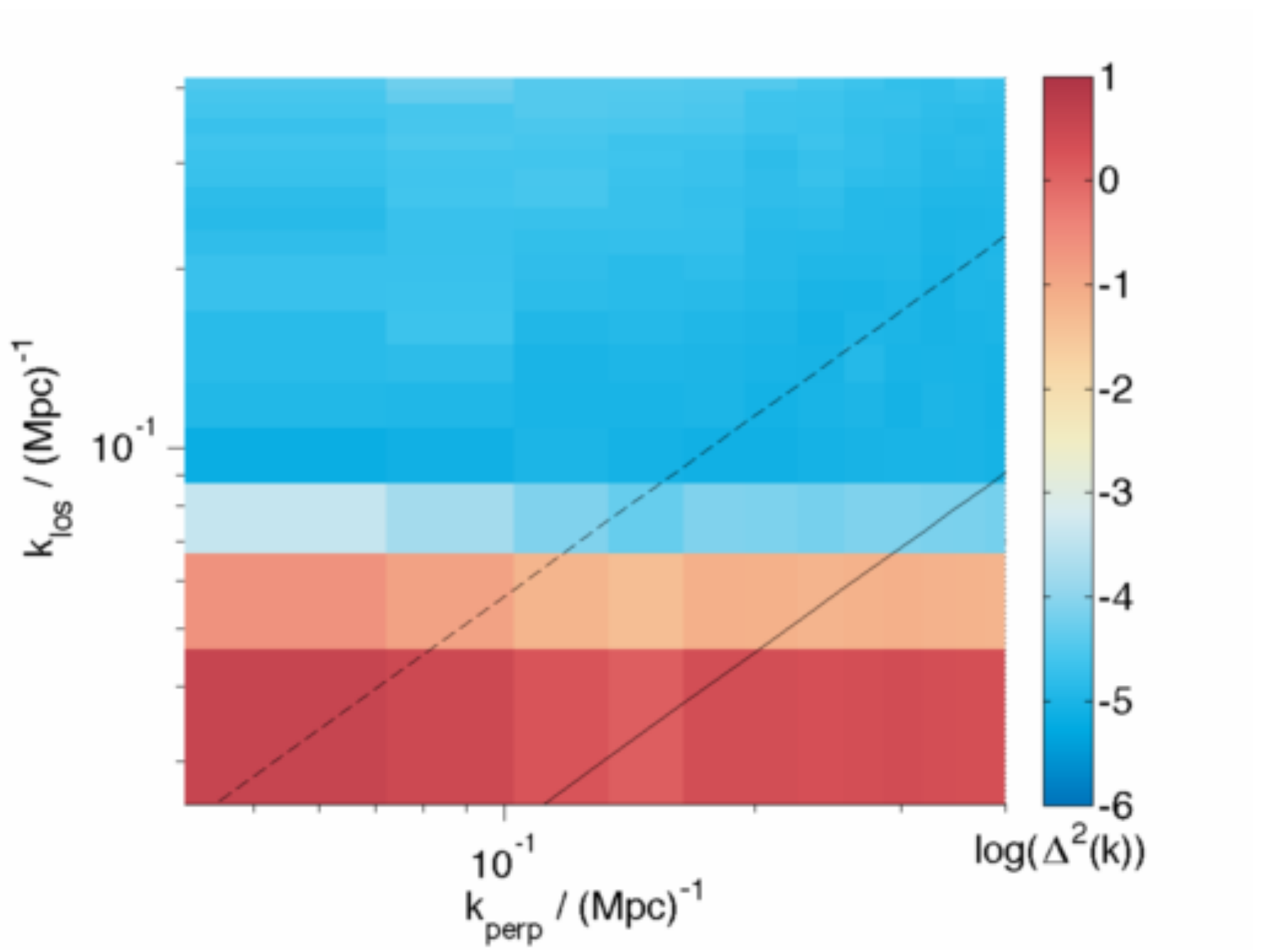}
\includegraphics[width=80mm]{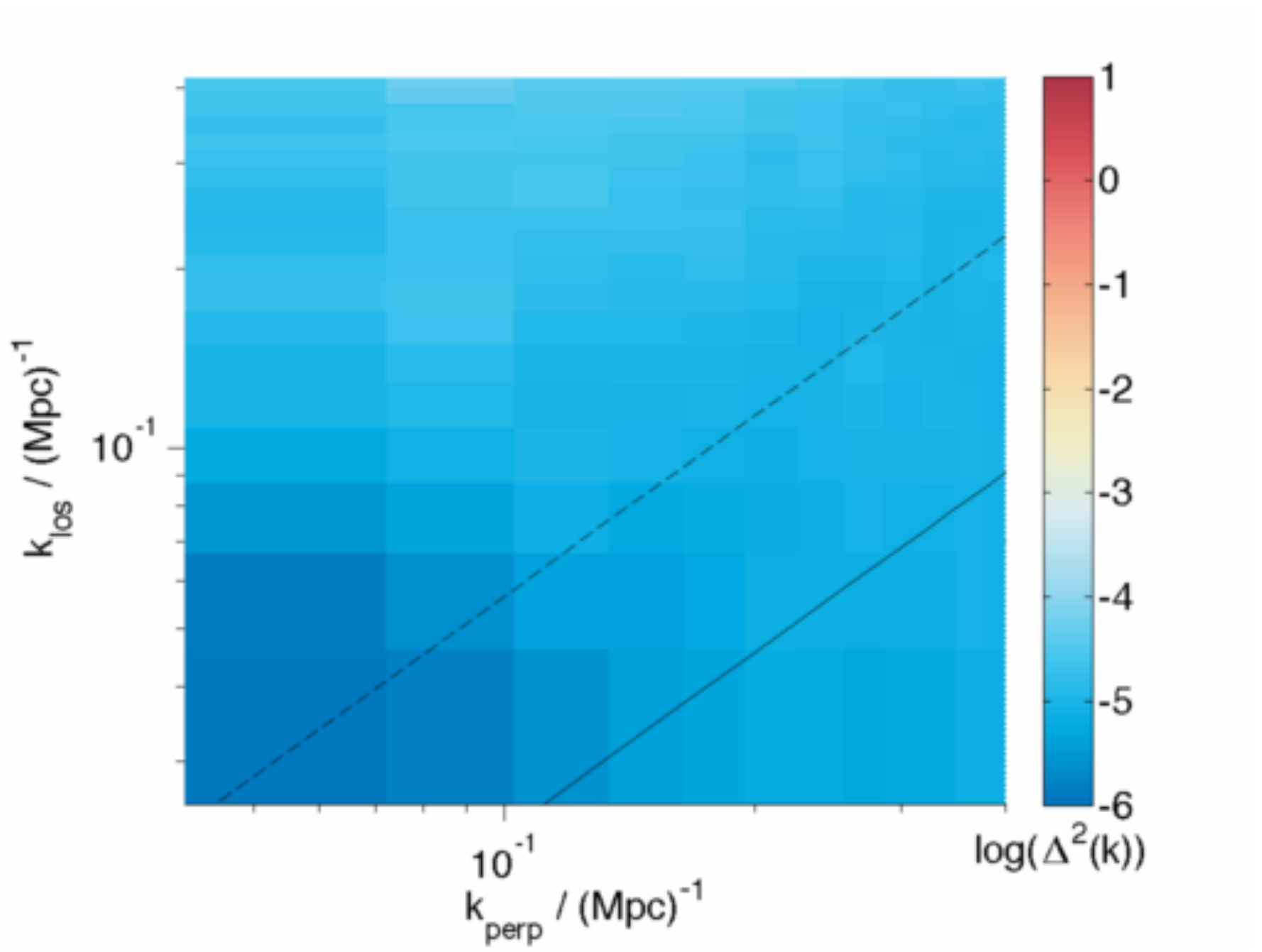}
\caption{The cylindrical power spectrum for the three scenarios, S1,
  S2 and S3 from top to bottom. In the left hand column is the total
  input signal for each scenario (i.e. the data for foreground avoidance) while on the right hand is the
  residual as produced by GMCA for each scenario (i.e. the data after
  foreground removal). Linestyles are as in Fig. \ref{win_comp}.}
\label{all_win}
\end{center}
\end{minipage}
\end{figure*}

We can take a first look at the power spectrum recovery of the two methods by looking at the
cylindrical power spectrum of Fig. \ref{all_win}. We do this for three scenarios. The first, S1, with the
LOFAR 600h noise as described above. The second, S2, we divide our noise
by 10 in an approximation to the expected SKA noise. Finally, we
look at the `perfect' situation where there is zero instrumental noise
in an effort to understand the foreground effects alone, S3. We can clearly see the dominant foreground
contribution in red at low $k_{los}$ for all three scenarios. For all scenarios we can see the contours which
were apparent in the cosmological signal plot.

In order to visualise the results of using the EoR window approach and
foreground removal we ran the
total input signal cube through GMCA assuming four components in the
foreground model. In the remaining three panels of Fig. \ref{all_win} we present the cylindrical
power spectrum of the three residuals cubes. It is clear that GMCA is able to remove the foreground contamination very well, though whether it affects the cosmological signal as a result of inaccurate foreground modelling is less clear.

\begin{figure*}
\begin{minipage}{170mm}
\begin{center}
\includegraphics[width=80mm]{{./Figures/cyl_ps_ratio_cs_zeromean_highres_115_000MHz_199_500MHz_4deg_K_exBN1D_han2D_fg_zeromean_highres_115_000MHz_199_500MHz_4deg_K_exBN1D_han2D_165_00MHz}.pdf}
\includegraphics[width=80mm]{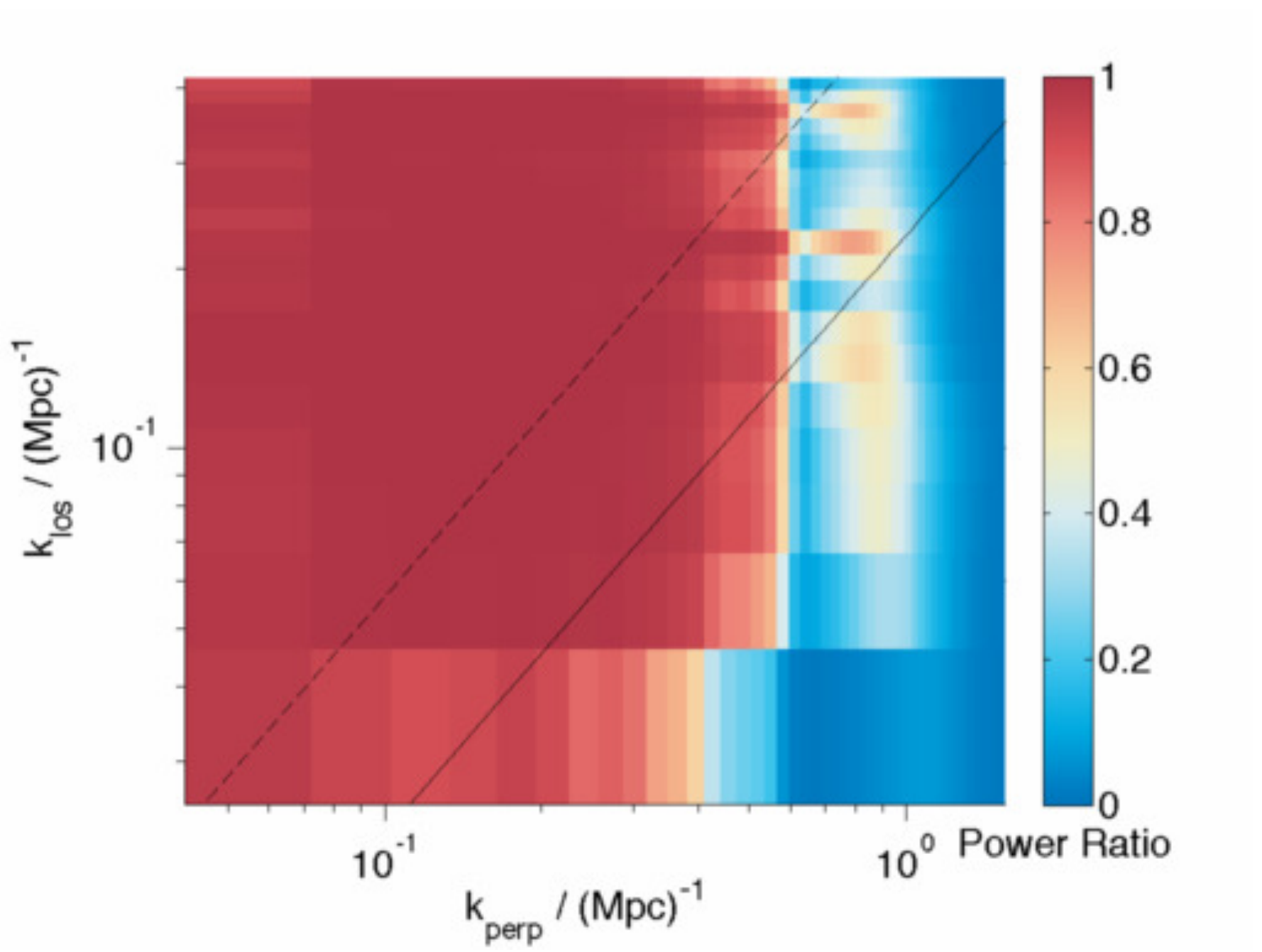}
\includegraphics[width=80mm]{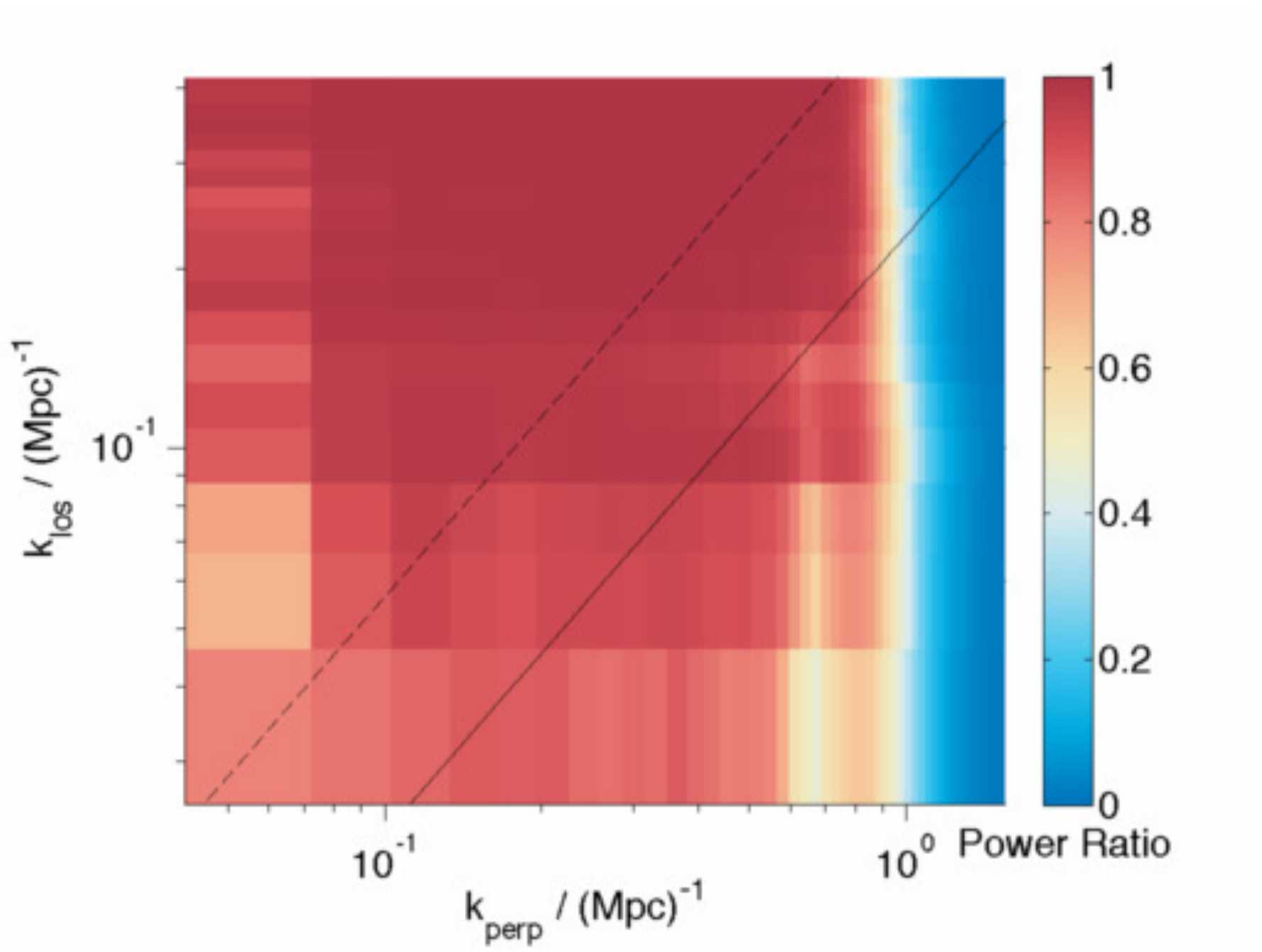}
\includegraphics[width=80mm]{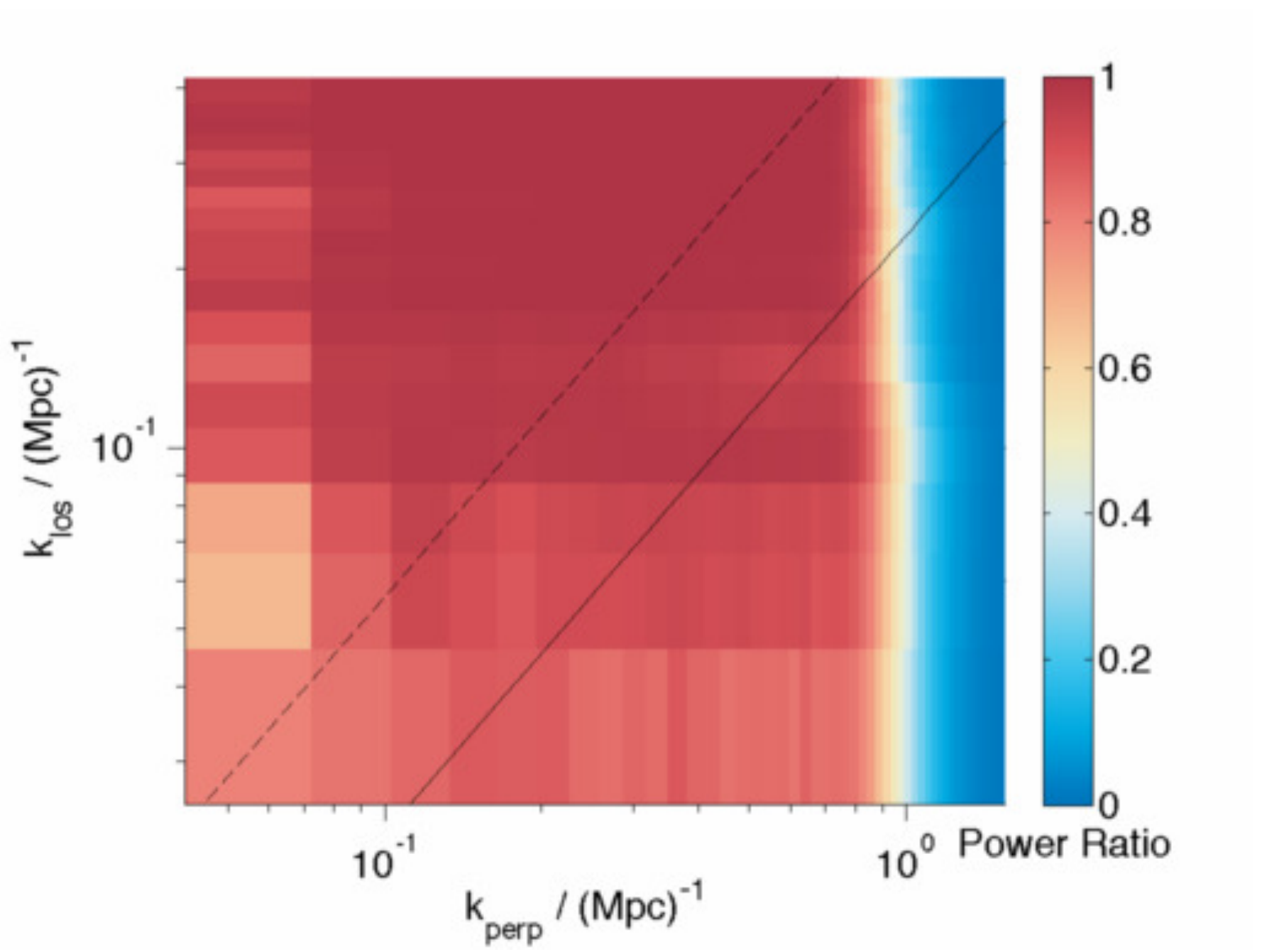}
\caption{In reading order: The ratio $\mathrm{cs}/(\mathrm{fg+cs})$ which relates to the foreground avoidance method assuming perfect noise knowledge. The ratios $\mathrm{cs}/(\mathrm{fg}_{\mathrm{fiterr}}+\mathrm{cs})$ for the three scenarios, S1,
  S2 and S3, which relate to the foreground removal method assuming perfect noise knowledge. If the ratio plot is equal to one then
  foreground contamination is nil. Linestyles are as in Fig. \ref{win_comp}.}
\label{ratios}
\end{center}
\end{minipage}
\end{figure*}

In order to assess the accuracy of the foreground removal
across the $k$-plane we calculate the foreground fitting errors which are the difference between the simulated foregrounds and those modelled by GMCA. We display the ratio
$\frac{cs}{\mathrm{cs}+\mathrm{fg_{fitterr}}}$ for each scenario in Fig. \ref{ratios}.  While, as stated before, direct comparison between foreground removal and foreground avoidance on data without action of a frequency dependent PSF should be treated with some caution, we can see how the methods work differently. We first of all see that with expected LOFAR noise we are able to recover a much greater portion of the window at $k_{los} < 0.09$ i.e. GMCA is successfully removing the foregrounds. We note however that compared to foreground avoidance less of the window is recovered at $k_{perp} > 0.4 \mathrm{Mpc}^{-1}$. This is due to the noise confusing the GMCA method and resulting in a less accurate foreground model. As we reduce the noise in S2 we see a similar range of $k_{perp}$ is recovered as foreground avoidance and for the ideal no-noise S3 we have almost the entire window recovered, barring the PSF action at $k_{perp} > 0.6 \mathrm{Mpc}^{-1}$. This is a most promising result for both methods as we see large portions of the window recovered. For current generation telescopes it may be that a combination of the methods proves most fruitful in order to access as much of the window as possible. If the desire is for as much good quality data as possible irrespective of scale, foreground avoidance will be useful. On the other hand, to access those smallest $k_{los}$ scales, foreground removal will be necessary. If the action of a frequency dependence PSF can be successfully mitigated then by the time the next generation data is available it seems that foreground avoidance will provide little advantage over foreground removal.

\begin{figure}
\begin{minipage}{70mm}
\begin{center}
\includegraphics[width=80mm]{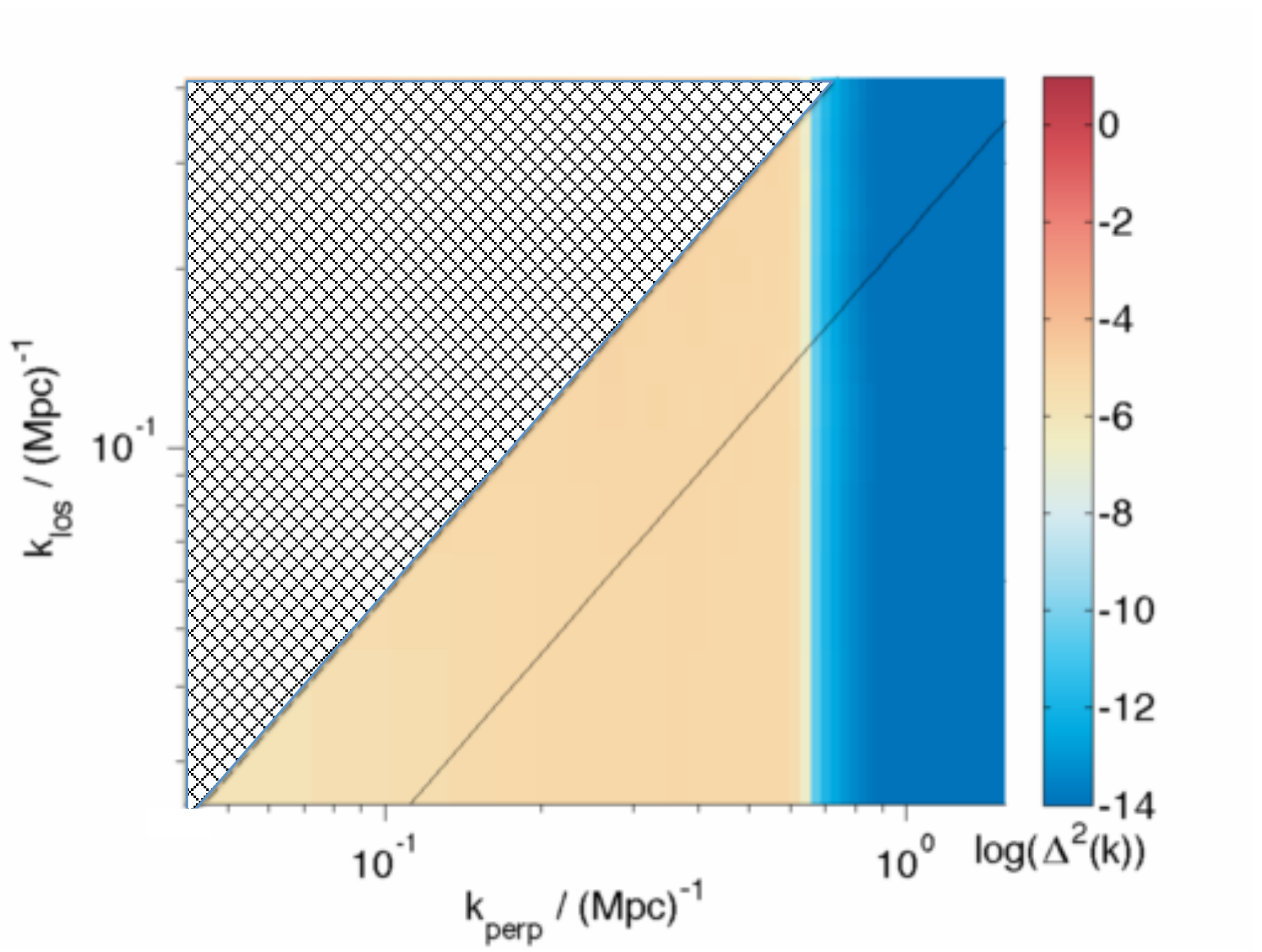}
\caption{A cartoon to show how we define the recoverable EoR window in our quantative calculation. The hashed area represents the total EoR window ($W_{FoV}$) which we consider recoverable if using the field of view wedge line. If all cells in this area were successfully recovered, we would deem the proportion of the window recovered to be 100\%. $W_{Beam}$ is defined as the region extending to the solid wedge line.}
\label{window_def}
\end{center}
\end{minipage}
\end{figure}

We can now make an attempt to quantify the amount of signal recovered. For an EoR window defined by the four instrumental boundaries and the wedge line we can ask how much of the signal to noise ratio within that window is recovered. We define this using a signal-to-noise ratio over all $k$ bins over the area of the EoR window extending either down to the beam wedge line or the field of view wedge line as defined by the hashed section in Fig. \ref{window_def}. We term these two windows $W_{Beam}$ and $W_{FoV}$ respectively, where $W_{Beam} > W_{FoV}$ in terms of area. We define the signal-to-noise ratio of the foreground removal methods as follows:

\begin{equation}
SN_{rem} = \frac{\sum_{ij}{P_{cs}(k_{ij})}}{\sum_{ij}{(P_{fg_{fiterr}}(k_{ij})+P_{no}(k_{ij}))}}
\label{SNrem}
\end{equation}

For the SN of foreground avoidance we add the condition that only $k$ bins with $k_{los}>0.09$ are included in the sums, and include the power spectrum of the foregrounds instead of the foreground fitting errors:

\begin{equation}
SN_{avoid} = \frac{\sum_{ij}{P_{cs}(k_{ij})}}{\sum_{ij}{(P_{fg}(k_{ij})+P_{no}(k_{ij}))}}
\label{SNav}
\end{equation}

\begin{figure}
\begin{minipage}{70mm}
\begin{center}
\includegraphics[width=80mm]{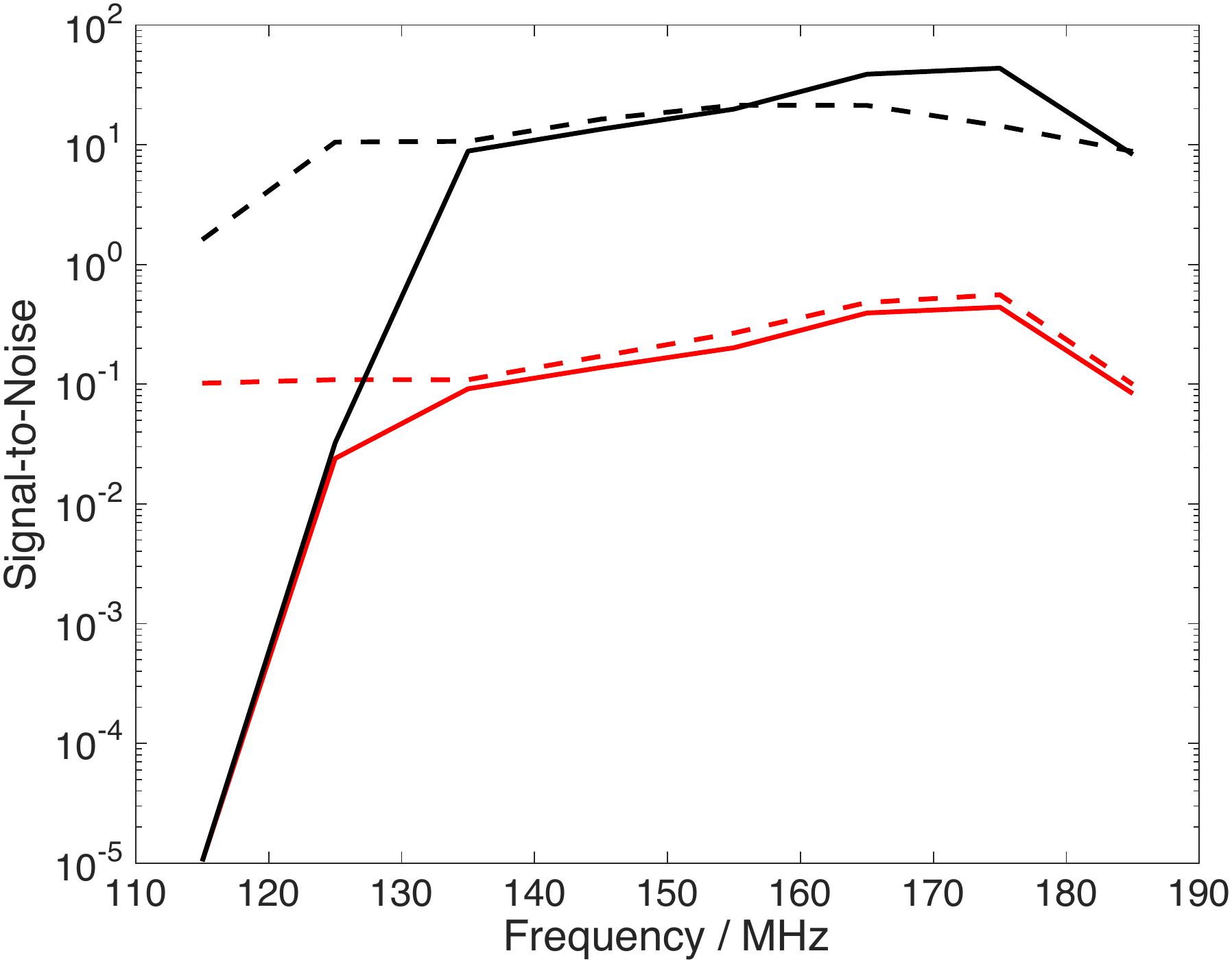}
\caption{The signal-to-noise ratio as defined in Equations \ref{SNav} and \ref{SNrem} as a function of central frequency in a 10 MHz bandwidth. The thick lines are for the $W_{FoV}$ definition of the window and the thin lines are for the $W_{Beam}$ definition.} 
\label{fig_SN}
\end{center}
\end{minipage}
\end{figure}

In Fig \ref{fig_SN} we see that the level of signal-to-noise recovered is dependent both on frequency and on the level of noise, as expected. We see that for the LOFAR scenario (red lines), foreground removal results in a slightly higher signal-to-noise ratio across the frequency range. In the lowest frequency bins the signal-to-noise ratio decreases significantly for foreground avoidance. This is because the cosmological signal decreases significantly at these frequencies and so the signal-to-foreground level is severely diminished. In contrast, foreground removal methods are able to remove the foregrounds in order to recover an improved signal-to-noise even very low signal-to-foreground levels. For the SKA scenario (black lines) we see the same trends, except that above 160 MHz, the signal-to-noise for foreground removal decreases slightly below the foreground avoidance line. This suggests that the foreground fitting errors are the dominate source of noise on the signal at these frequencies - not the noise or foregrounds themselves. This suggests that for the next generation of telescopes further optimisation of the foreground removal codes may be necessary. As pointed out before, a direct comparison is not completely robust without proper frequency dependence of the PSF and the signal-to-noise does not take into account that if one were to desire only scales between 0.4 < $k_{perp}$ < 0.6 for example, then it would be best to choose foreground avoidance even for LOFAR levels of noise.

\begin{figure*}
\begin{minipage}{170mm}
\begin{center}
\includegraphics[width=80mm]{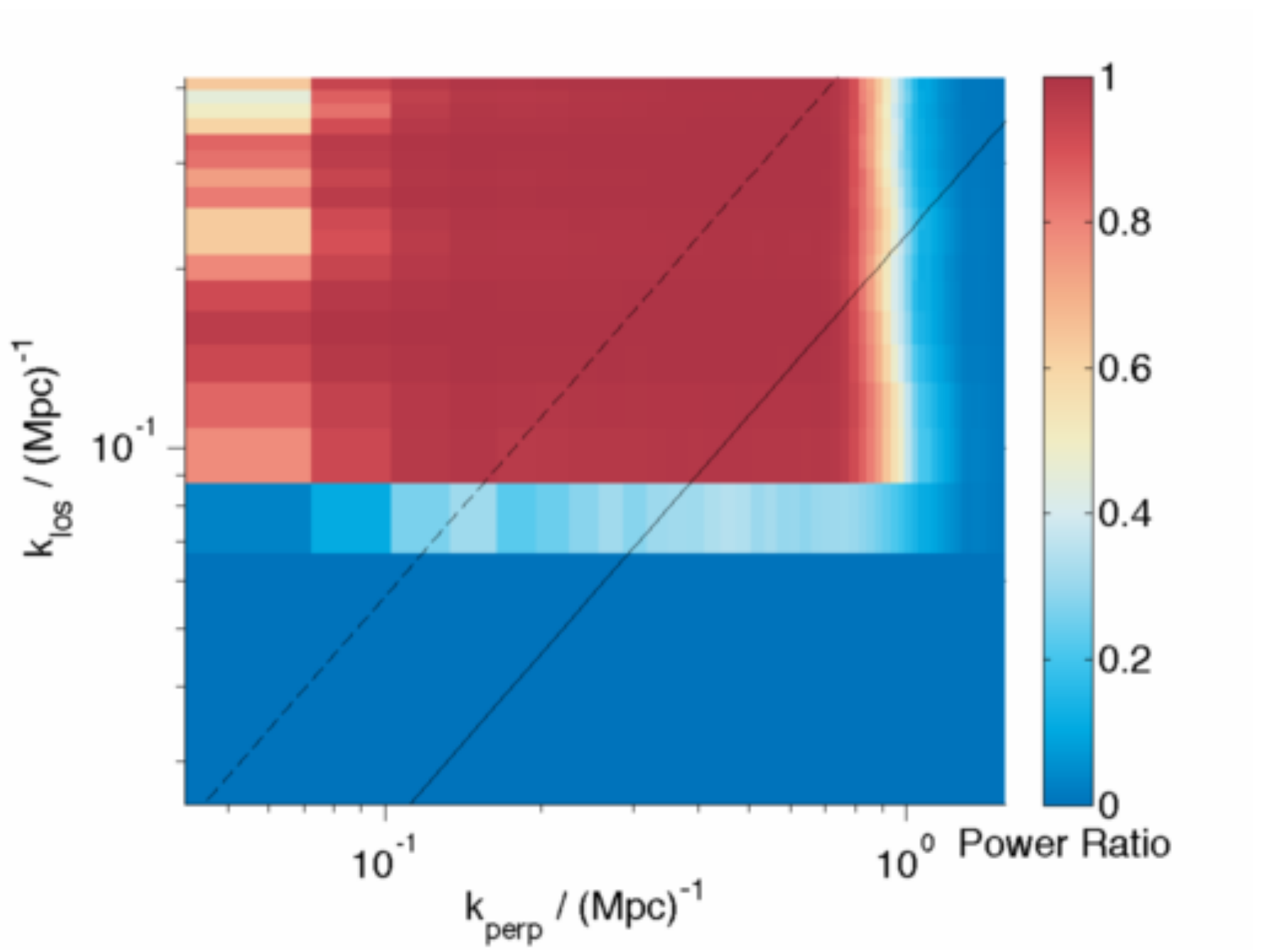}
\includegraphics[width=80mm]{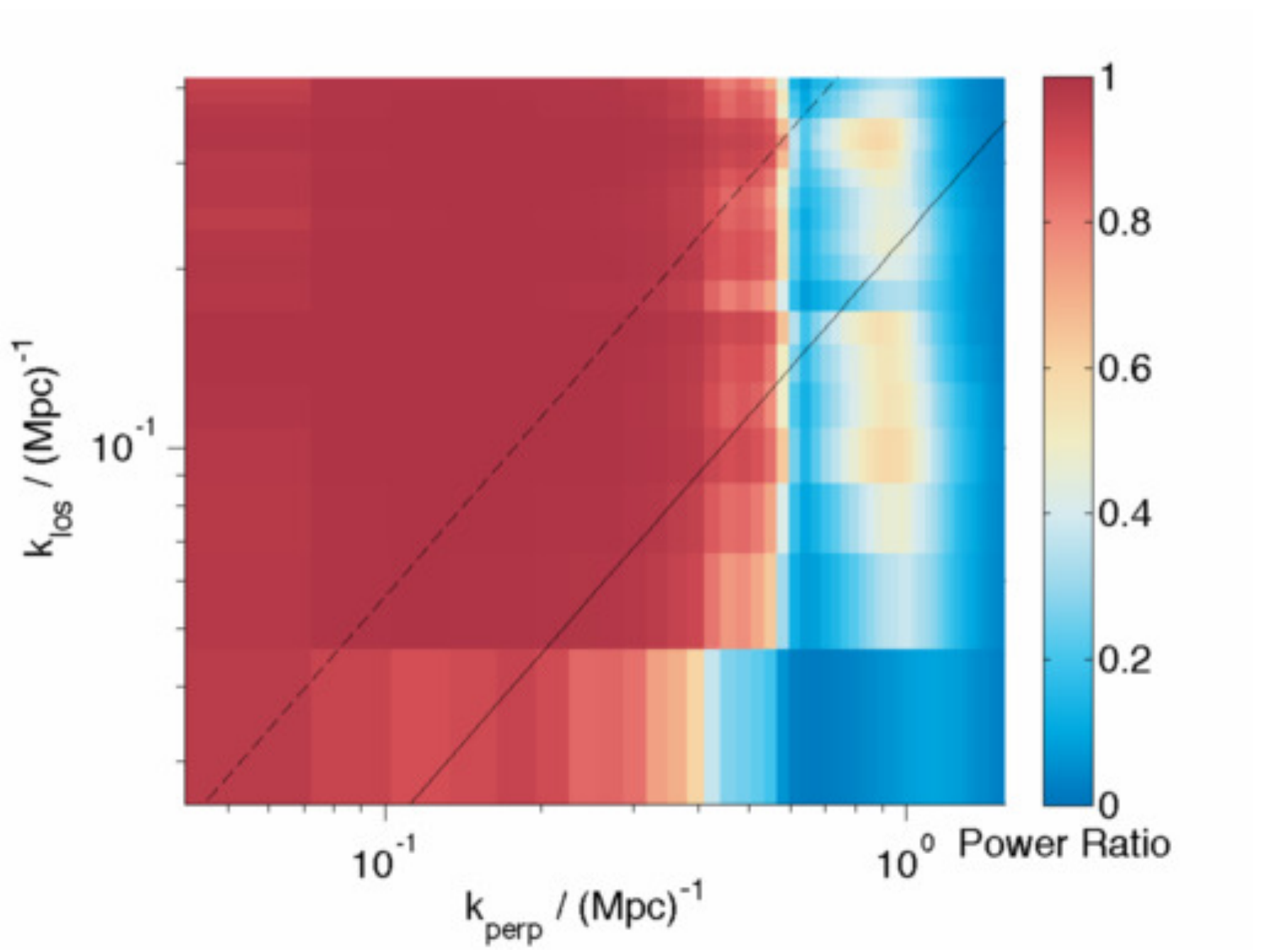}
\includegraphics[width=80mm]{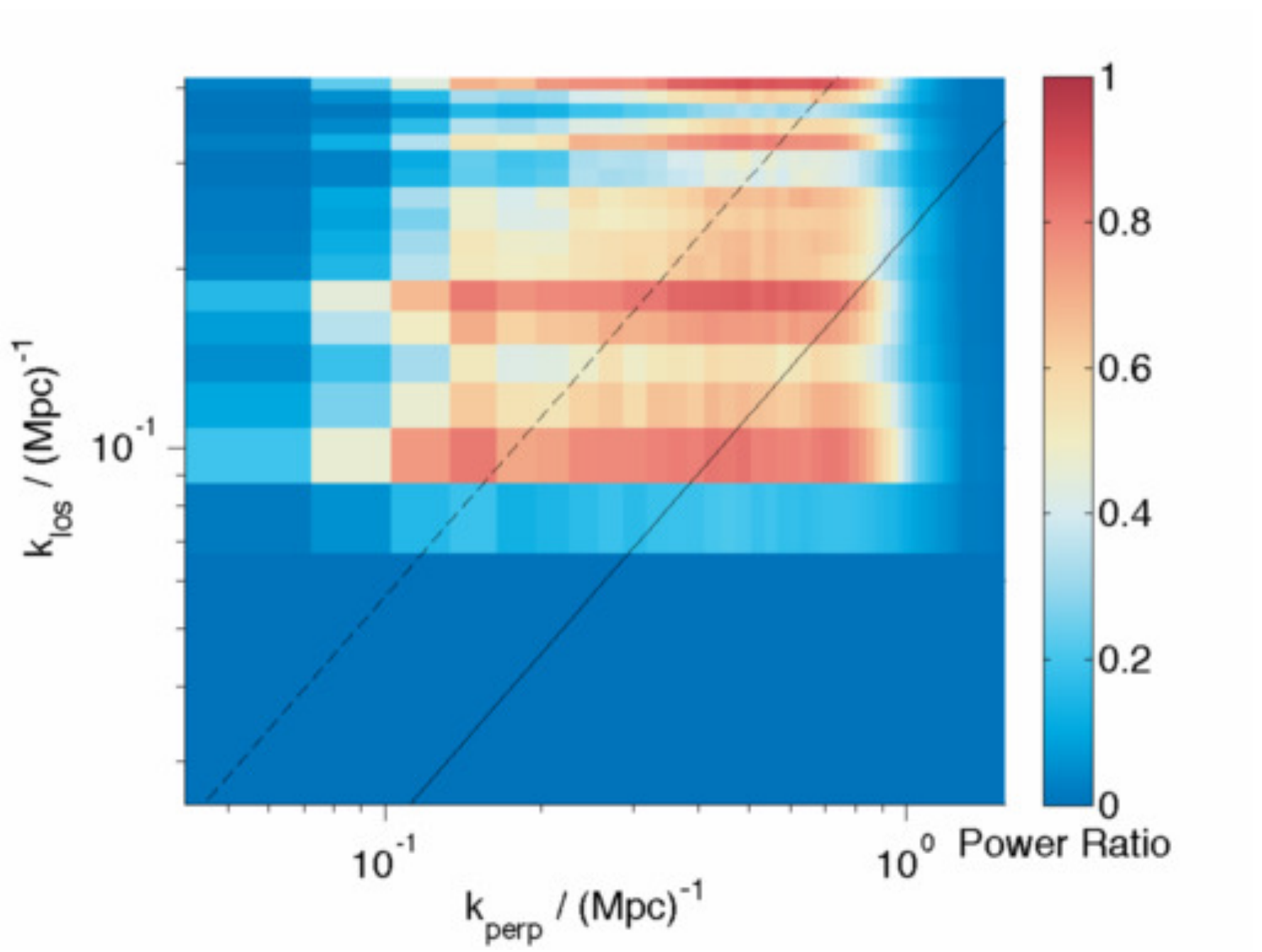}
\includegraphics[width=80mm]{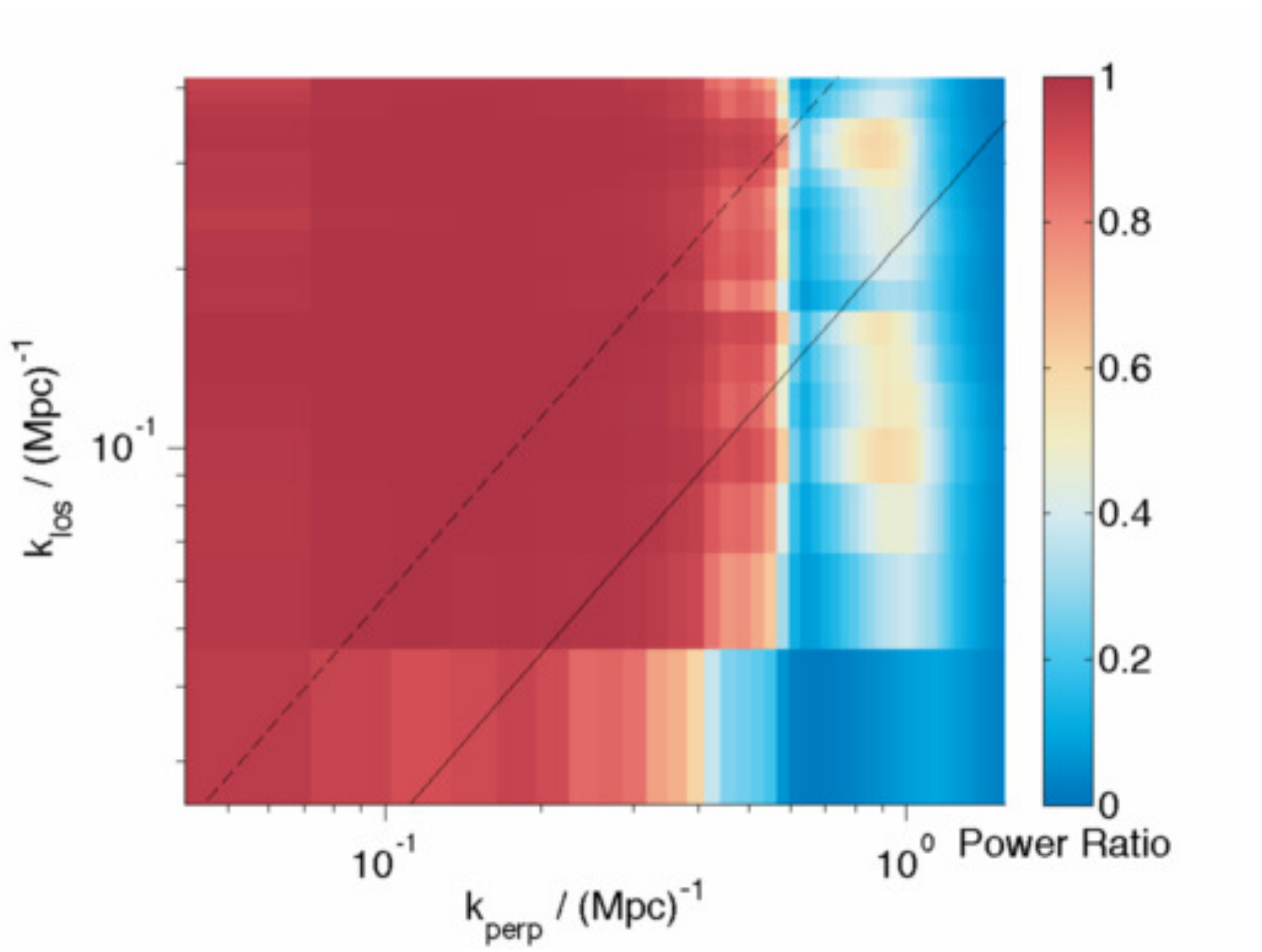}

\caption{In the left column we show the ratio cs/cs+fg for the $0.1\%$ and $1\%$ LOS varying foreground models (top and bottom respectively). In the right column we show the same ratio but with the foreground fitting errors as a result of performing GMCA on the new foreground cubes with LOFAR noise. We see that foreground avoidance is affected badly by introducing a $1\%$ LOS variation, while GMCA is still able to make a good recovery, under the assumption of common resolution channels}. Linestyles are as in Fig. \ref{win_comp}.
\label{fg_models_1}
\end{center}
\end{minipage}
\end{figure*}

\begin{figure*}
\begin{minipage}{170mm}
\begin{center}

\includegraphics[width=80mm]{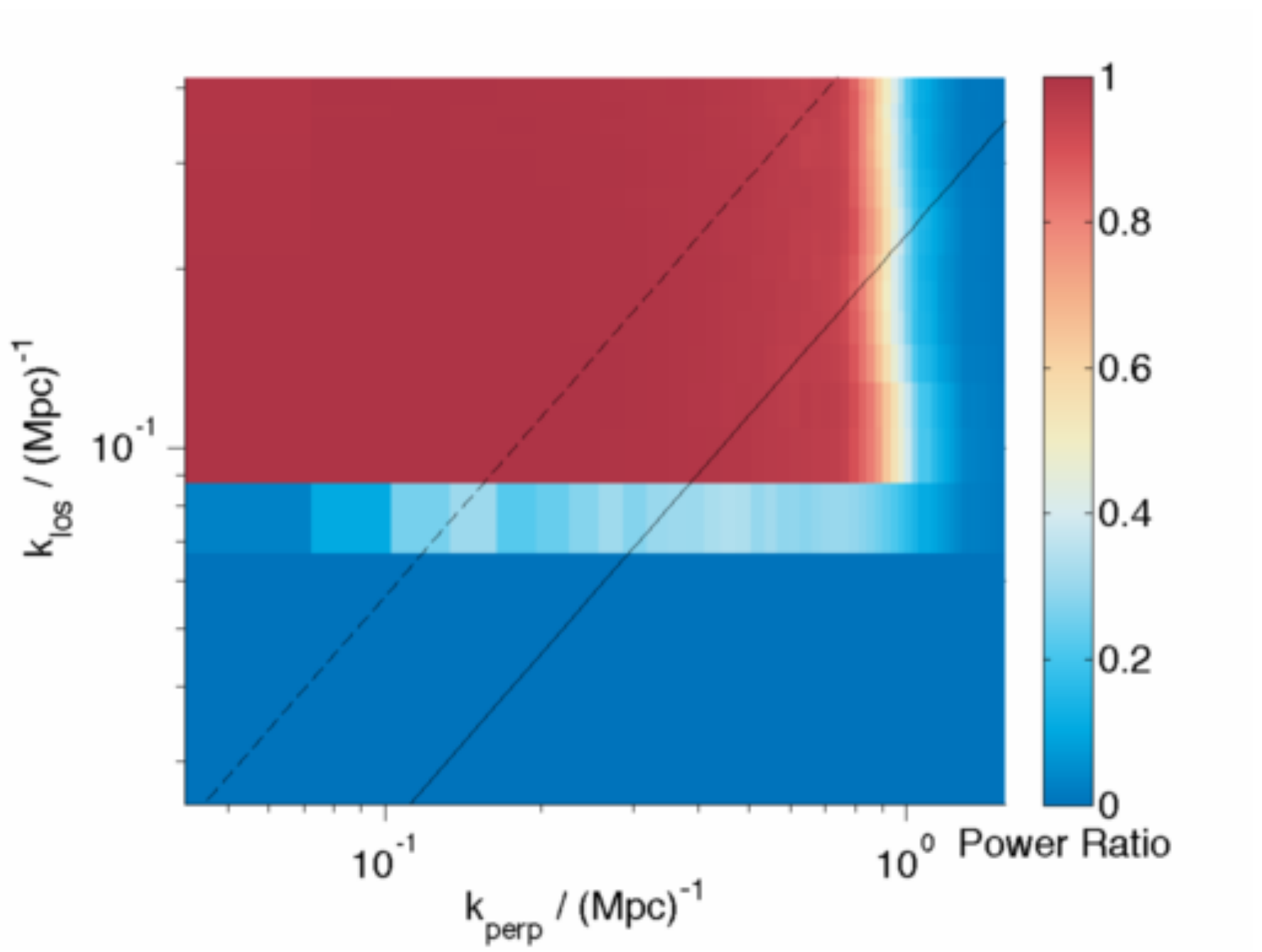}
\includegraphics[width=80mm]{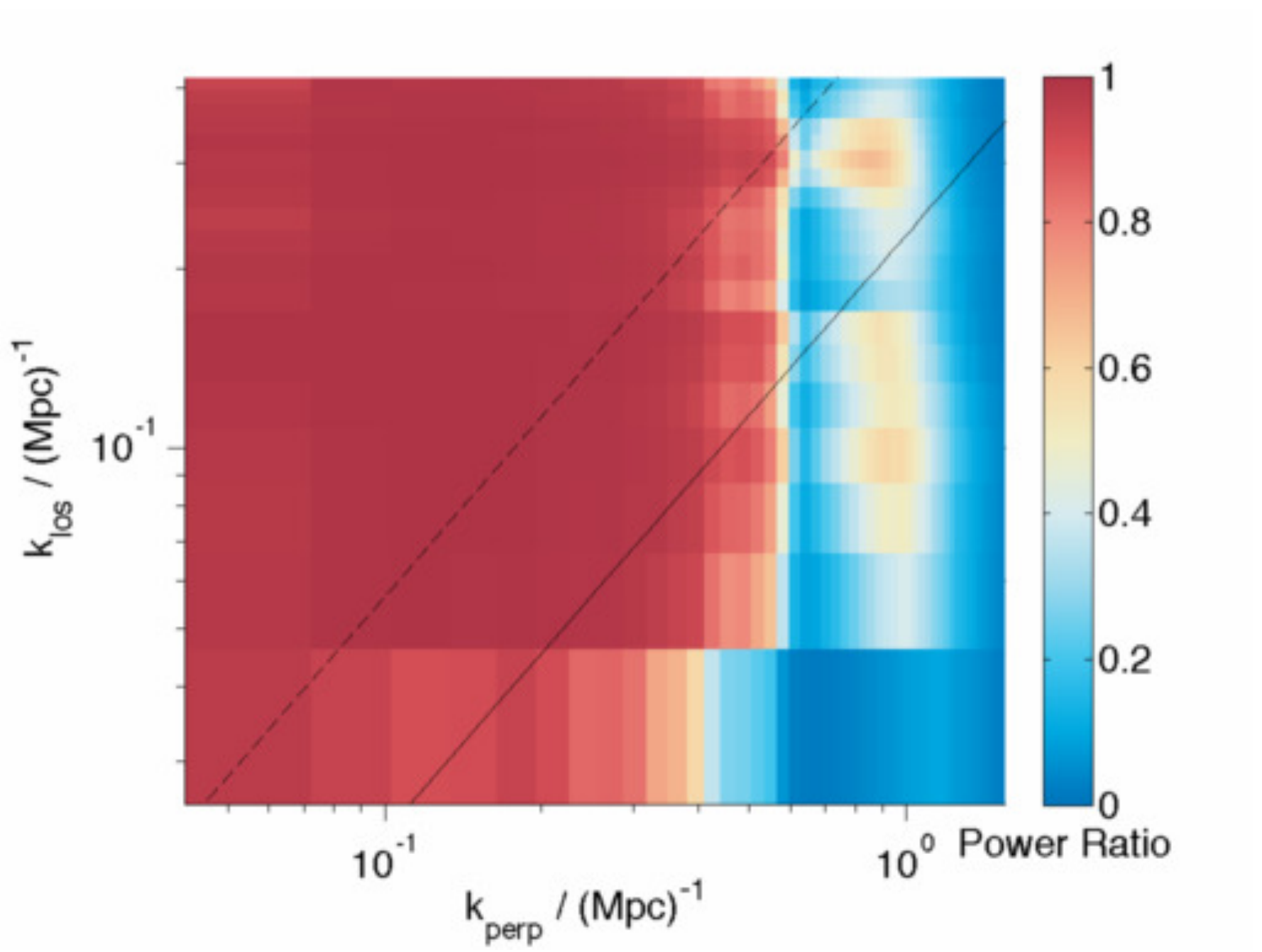}
\includegraphics[width=80mm]{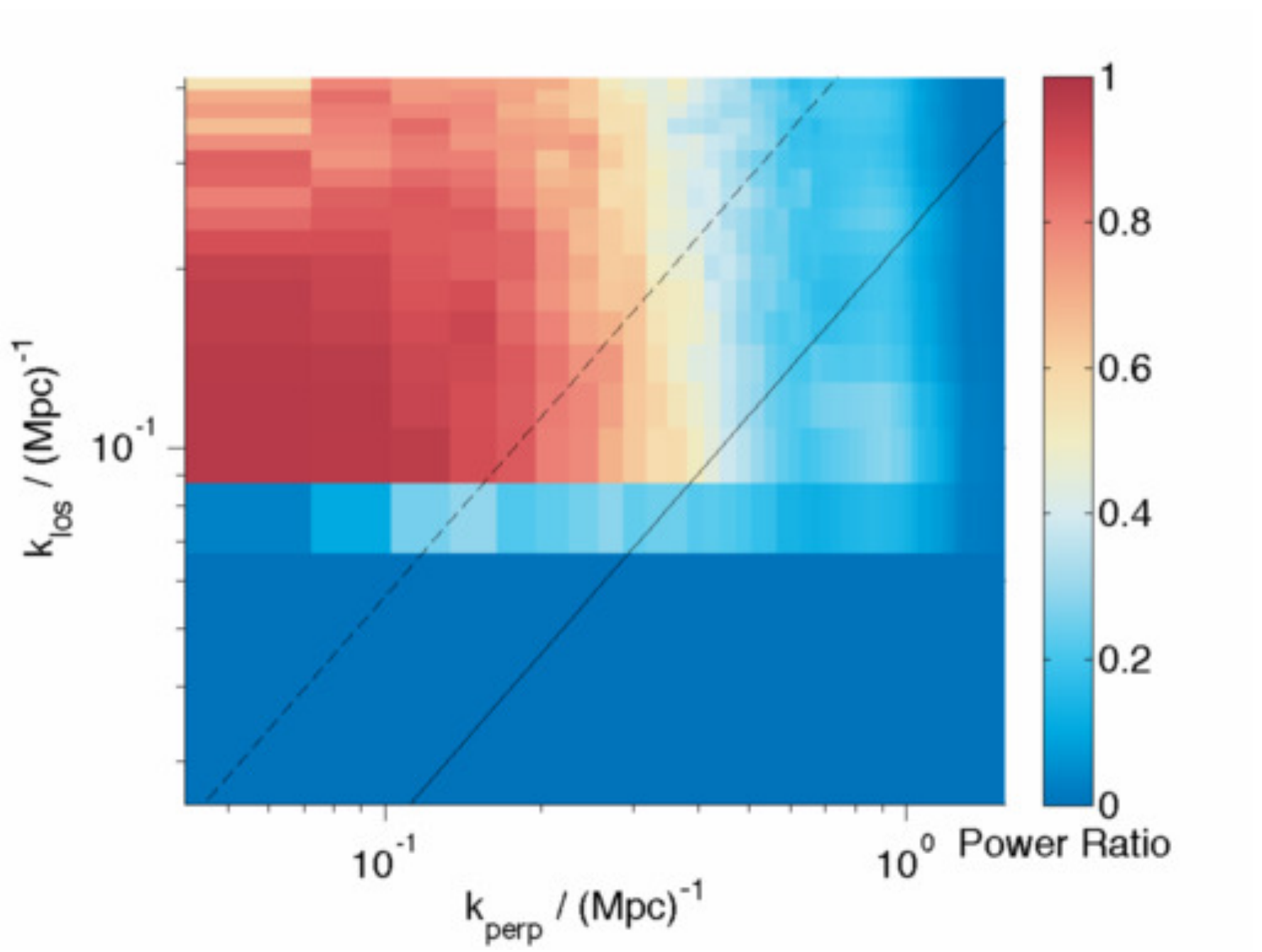}
\includegraphics[width=80mm]{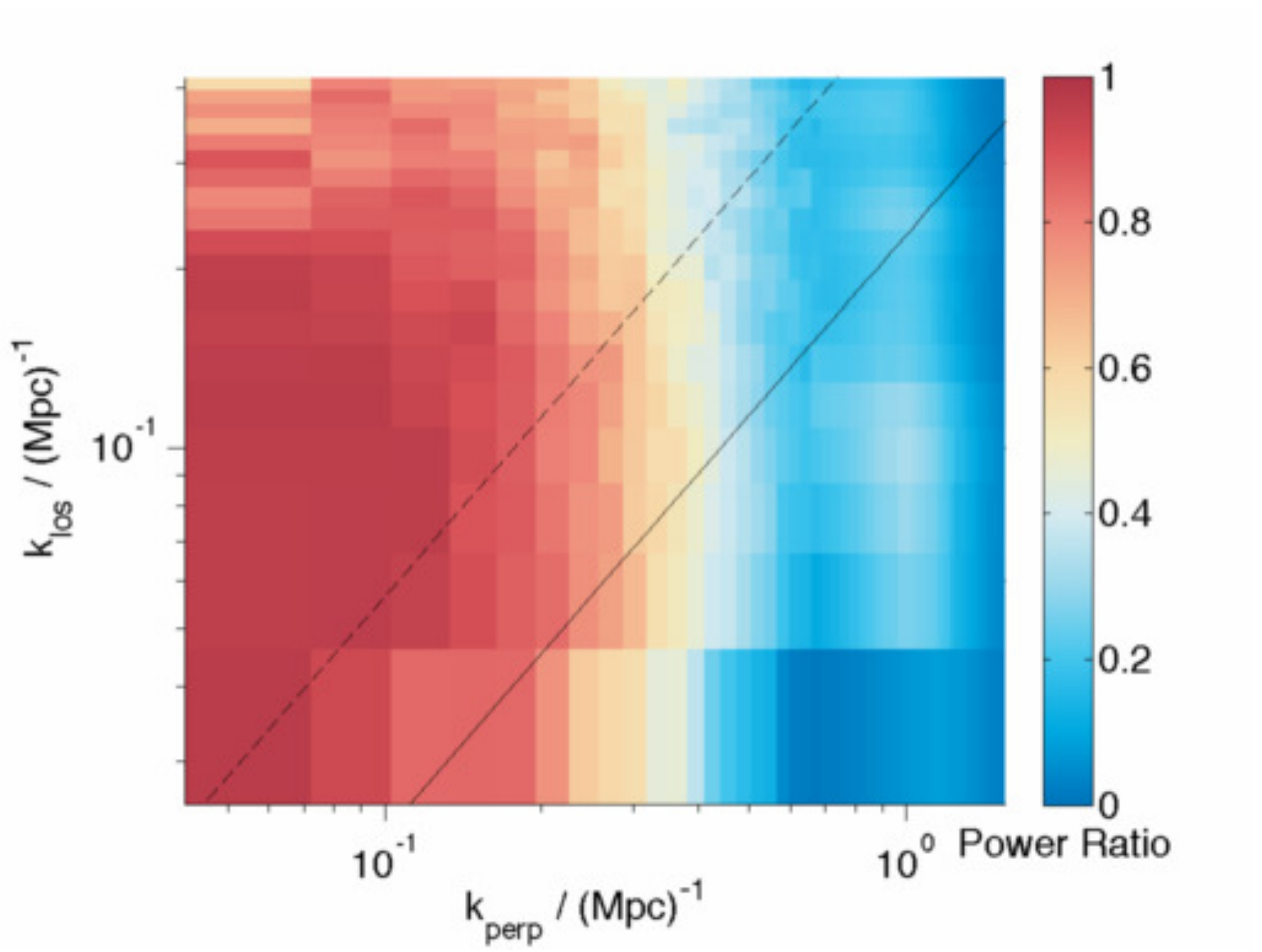}

\caption{In the left column we show the ratio cs/cs+fg for the $0.1\%$ and $1\%$ (top and bottom respectively) LOS+spatial varying foreground models. In the right column we show the same ratio but with the foreground fitting errors as a result of performing GMCA on the new foreground cubes with LOFAR noise. While GMCA is able to cope fairly well with a $0.1\%$ variation, the $1\%$ LOS+spatial variation results in a dramatic reduction in recovery across the $k$ range. Linestyles are as in Fig. \ref{win_comp}.}
\label{fg_models_2}
\end{center}
\end{minipage}
\end{figure*}

\subsection{Less Smooth Foreground Models}

The assumption of smooth foregrounds such as those modelled in this paper is key to the success of many parametric and some non-parametric (e.g. \citet{harker09b}) foreground removal methods. In this section we assess the degree to which foreground avoidance and removal can be compromised by relaxing this assumption. We investigate four possibilities. Firstly, we adjust each slice of our clean foreground simulation by multiplying with a random number drawn from a Gaussian distribution with standard deviation of 0.01. This very roughly models an inherent 1\% variation of the foreground magnitude along the line-of-sight. We then do the same but with a standard deviation of 0.001 to create a 0.1\% variation scenario. We see the results of applying the methods to total signal cubes made with these adjusted foreground cubes in Fig. \ref{fg_models_1}. We run GMCA on the data cubes including LOFAR-like instrumental noise and so the resulting ratios can be directly compared to the top panels of Fig. \ref{ratios}. We see that neither cube is too adversely affected by the 0.1\% variation, though the foreground avoidance method does suffer some degradation at $k_{perp} < 0.2 \mathrm{Mpc}^{-1}$ due to the small-scale wiggle introducing structure into the foregrounds which cannot be confined purely in a band at low $k_{los}$. However for 1\% variation we see a marked difference. While GMCA is still able to model to foregrounds quite accurately for $k_{perp} < 0.6 \mathrm{Mpc}^{-1}$, the EoR window for foreground avoidance is much obscured. This is an important drawback of foreground avoidance to note. While there is little argument that the physical processes producing much of the foregrounds result in a smooth frequency dependence, it is not clear how the instrumentation and data reduction pipeline might encroach on this smoothness. Blind methods such as GMCA provide a clear advantage in such a case, as no assumption of smoothness is made. In contrast, for foreground avoidance to work it must assume the foregrounds are confined to within a clear low $k_{los}$ area. If in the case of a significant LOS variation this assumption fails and the foreground contamination within the EoR window is too high. Note that we present the "best-case" for foreground removal here - where the frequency dependence of the PSF has been perfectly mitigated.

Next we make the adjustment to the foregrounds spatially varying by multiplying every pixel by a random number as described above. This is in addition to the LOS variation. This emulates some form of instrumental calibration error such as leakage of polarized foreground from the Stokes Q and U channels to the Stokes I channel. In Fig. \ref{fg_models_2} we see that even with both spatial and LOS $0.1\%$ variation, there is no reduction in the quality of recovery for either method. However once this variation is increased to $1\%$ GMCA shows a marked decrease in accuracy in its foreground model and recovery across the whole $k$ range, though especially at $k_{perp} > 0.3$, where the small scale spatial wiggle is at its most significant. This is intriguing as it gives us insight into how GMCA uses sparsity to define the foreground components. While the smoothness of the foregrounds may aid GMCA in finding a basis in which a foreground component can be considered sparse, it is apparent from these results that it is not this smoothness which is essential for the method to work. However, once the spatial variation is included, GMCA fails to model the foreground accurately, indicating that it is the spatial correlation of the foregrounds which enables GMCA to find a sparse description of the foreground signal.
Interestingly foreground avoidance actually seems to do a better job than with LOS variation alone. We believe that this is because the introduction of a spatial variation softens the variation along the line of sight somewhat, confining the foregrounds once again to a low $k_{los}$ area. This could be a quirk of our particular simulations of the variations however and may warrant further investigation.

We note that the foreground models here are not an attempt to model specific instrumental or physical effects. We intend to explore a fully physically-motivated foreground leakage model in further work and simply present the toy models here as a first step to understanding the sensitivity of the methods to non-smooth foreground models. To truly assess the ability for GMCA to model the extra degrees of freedom introduced by a non-smooth foreground model would require a Bayesian model selection algorithm in order to select for the model with the most likely number of foreground components. The results presented here are in fact a non-optimal presentation of GMCA in this respect as we have not changed the number of components used by GMCA to model the non-smooth foregrounds. We leave the development of a Bayesian model selection algorithm and resulting analysis to further work.

\section{Conclusions}
\label{conc}

This paper set out to consider the loss of sensitivity in the power spectrum recovery using foreground avoidance, while also presenting how a foreground removal method currently adopted in EoR pipeline preserved that sensitivity in the optimal case of common resolution channels. This is a timely investigation due to the current data analysis being carried out by several major radio telescope teams around the globe in order to uncover the cosmological signal for the first time. Being a first detection of a rather unconstrained entity, we need to be confident in our methods for recovering it and understand how different methods might produce different results.

We aimed to begin to understand this by applying both techniques to three sets of simulations: one with LOFAR level noise, one with SKA level noise and one with no noise at all. The comparison was an extremely fruitful one and very promising with respects to both methods. We did however ascertain several differences between the methods:

\begin{itemize}
\item While the omission of low $k_{los}$ scales in foreground avoidance undoubtedly prevents foreground removal bias within the remaining signal, the amount of cosmological signal at those same omitted scales is not negligible. This could have a serious impact on investigations of, for example, redshift space distortions, as pointed out in \citet{pober14}.
\item Pursuing foreground removal results in a more complete
  reconstruction at low $k_{los}$ however there is a worse
  recovery at $k_{perp} > 0.6 \mathrm{Mpc}^{-1} $ due to foreground fitting
  errors. This is due to the noise confusing GMCA. The recovery improves as we go to lower noise scenarios, with the no-noise scenario recovering the same range of $k_{perp}$ scales as foreground avoidance. 
\item Quantifying the amount of signal recovered in the EoR window, we find that foreground removal recovers a greater signal-to-noise ratio than foreground avoidance across the frequency range for the LOFAR scenario. While for the SKA scenario, both methods recover a greater signal-to-noise, as expected, foreground removal does appear to overfit the foregrounds at the higher frequencies when all scales are considered and this needs to be investigated further. It may be that at SKA noise levels the number of components on the foreground model needs to be more carefully chosen to avoid over-fitting. 
\item Neither foreground avoidance or foreground removal is too adversely affected by a LOS or spatial variation of 0.1\%. 
\item When a LOS-only variation of $1\%$ is introduced we see that while GMCA can still recover the cosmological signal to the same degree, the EoR window is far too obscured for foreground avoidance to be an effective method.
\item For a spatial and LOS variation of 1\%, GMCA is diminished in its accuracy however can still recover a reasonable area at low $k_{perp}$ and still a larger area than foreground avoidance. Interestingly foreground avoidance seems to do better with spatial and LOS variation than with the LOS variation alone. We believe this to be a as a result of the spatial variation diminishing the LOS variation by chance.

\item At current generation noise levels it would probably be advantageous to use both methods in order to recover as much as the window as possible. However once we reach next generation noise levels the advantage of using foreground avoidance is less clear assuming the satisfactory mitigation of the frequency dependent PSF. 

This was a basic first look at the loss of sensitivity invoked by both methods and there is still much work to be done. The frequency dependence of the PSF is a known issue with respects to the cleanliness of the EoR window above the wedge due to mode-mixing. While for this first attempt we enforced a common resolution motivated by the recent results of a successful weighting scheme, we will need to consider how this frequency dependence can be mitigated by both methods in order to provide an equal footing comparison. We also would like to consider the wedge in more detail by including bright point sources in our analysis and modeling their inaccurate removal. 

\end{itemize}

\section{Acknowledgments}
EC acknowledges the support of the Royal Astronomical Society via a RAS Research Fellowship. EC would like to thank Ajinkya Patil, Cathryn Trott, Jonathan Pritchard, Catherine Watkinson, Suman Majumdar, Geraint Harker and Harish Vedantham for useful discussions. FBA acknowledges the support of the Royal Society via a University Research Fellowship. VJ would like to thank the Netherlands Foundation for Scientific Research (NWO) for financial support through VENI grant 639.041.336.
The work involving OSKAR was performed using the Darwin Supercomputer of the University of Cambridge High Performance Computing Service (http://www.hpc.cam.ac.uk/), provided by Dell Inc. using Strategic Research Infrastructure Funding from the Higher Education Funding Council for England and funding from the Science and Technology Facilities Council.

\bibliography{main_3.0}
\bibliographystyle{mn2e}  

\end{document}